# Perturbative Growth of Cosmological Clustering II: The Two Point Correlation.

Somnath Bharadwaj
Raman Research Institute, Bangalore 560 080, India. [1]
and
Joint Astronomy Program, Indian Institute of Science, Bangalore 560 012, India.
email somnath@rri.ernet.in

## ABSTRACT

We use the BBGKY hierarchy equations to calculate, perturbatively, the lowest order nonlinear correction to the two point correlation and the pair velocity for Gaussian initial conditions in a critical density matter dominated cosmological model. We compare our results with the results obtained using the hydrodynamic equations which neglect pressure and we find that the two match, indicating that thare are no effects of multistreaming at this order of perturbation. We analytically study the effect of small scales on the large scales by calculating the nonlinear correction for a Dirac delta function initial two point correlation. We find that the induced two point correlation has a $x^{-6}$ behaviour at large separations. We have considered a class of initial conditions where the initial power spectrum at small $k$ has the form $k^n$ with $0 < n \leq 3$ and have numerically calculated the nonlinear correction to the two point correlation, its average over a sphere and the pair velocity over a large dynamical range. We find that at small separations the effect of the nonlinear term is to enhance the clustering whereas at intermediate scales it can act to either increase or decrease the clustering. At large scales we find a simple formula which gives a very good fit for the nonlinear correction in terms of the initial function. This formula explicitly exhibits the influence of small scales on large scales and because of this coupling the perturbative treatment breaks down at large scales much before one would expect it to if the nonlinearity were local in real space. We physically interpret this formula in terms of a simple diffusion process. We have also investigated the case $n = 0$ and we find that it differs from the other cases in certain respects. We investigate a recently proposed scaling property of gravitational clustering and we find that the lowest order nonlinear terms cause deviations from the scaling relations which are strictly valid in the linear regime. The approximate validity of these relations in the nonlinear regime in N-body simulations cannot be understood at this order of evolution.

[1] Postal address

## 1. Introduction

In the first paper of this series (Bharadwaj 1994), which we shall refer to as part I, we presented a formalism for perturbatively studying the growth of cosmological correlations using the BBGKY hierarchy. We also appplied this formalism to the particular case when the initial disturbance was Gaussian and could be described by the two point correlation function, all higher correlations being zero. We used the formalism to calculate, in a universe with $\Omega = 1$, the induced three point correlation function to the lowest order of perturbation theory. We found that the three point correlation function at any length scale depends on the two point correlation at all smaller scales i.e. the small scales influence the large scales. This disgrees with the commonly used 'hierarchical form' where one assumes that the three point correlation at any length scale depends only on the two point correlation at the same scales.

The nonlinear evolution of the two point correlation function is driven by the three point correlation function through the tidal force of the third particle. In this paper we carry the perturbative scheme one step further and calculate the lowest order nonlinear correction to the two point correlation function. Before presenting this calculation we discuss some general issues about the scheme of calculation being used in this paper. We would also like to compare it with another scheme whose usage is more prevalent in the current literature.

We consider a universe filled with collisionless particles that are initially uniformly distributed and move according to Hubble's law. We study the statistical properties of disturbances in such an universe. To do this we construct an ensemble with each member corresponding to a different relization of the disturbance. We follow the evolution using Newtonian dynamics and use this to study the growth of correlation functions. The disturbances that are considered are such that initially they have a single streamed flow where all of the particles at any point have the same velocity and this velocity field is a smooth function of position. The initial mass density too is a smooth function . As the disturbance evolves the trajectories of particles intersect and the streams pass through one another interacting only through gravity. When this occurs the velocity and density are no longer smooth functions and they become singular and multi valued at points. This process is best illustrated by the Zeldovich approximation (Zeldovich 1970). Even though we start the evolution with a single streamed flow everywhere, multistreamed flows with particles with different velocites at the same point are inevitable. This fact forms the motive for the comparison of the two schemes of calculation given below.

The scheme of calculation used in this paper is based on using the equation of motion of the individual particles to obtain equations for the evolution of ensemble averaged phase space distribution functions (i.e. the BBGKY hierarchy). We consider velocity moments of the distribution functions and evolve them perturbatively. This gives the evolution of correlation functions. These equations are valid in both the single stream and multi stream regimes.

The scheme with which we would like to make the comparison evolves individual members of the ensemble using the hydrodynamic equations i.e. the continuity equation, Euler's equation and Poisson's equation. Pressure and other higher moments of velocity, which are initially zero, are assumed to be zero at later times and this scheme may be described as the single stream approximation. These equations become invalid once multistreaming occurs.

Although the two schemes are equivalent in the initial stages of the evolution when it is a single streamed flow, the single stream approximation breaks down as the evolution proceeds. Once this happens we expect the two schemes to give different results. The lowest order nonlinear correction to the two point correlation function has beenn calculated by Juszkewicz (1981), Vishniac ( 1983), and Makino, Sasaki and Suto (1992). Fry (1994) has calculated the higher order nonlinear corrections to the two point correlation function. All these calculations have been done using the single stream approximation.

In section 2 we use the BBGKY hierarchy to calculate the lowest order nonlinear correction to the two point correlation function and compare it with the results obtained in the single stream approximation. In section 3 we calculate the lowest order nonlinear correction to the pair velocity.

Unlike the linear evolution, the nonlinear evolution exhibits coupling of the various scales. This has been examined analytically by Suto and Sasaki (1991) and Makino et al. (1992). This issue has also been studied by Juszkewicz, Sonoda and Barrow (1984) and Hansel et. al. (1985), and for the particular case of the CDM initial condition it has been studied by Coles (1990), Jain and Bertschinger (1994), and Baugh and Efstathiou (1994). Most of the analysis has been in Fourier space. In this paper we address this question in real space. The coupling of the various scales in the nonlinear epoch has also been studied using N-body simulations by Mellot and Shandarin (1990), Little, Weinberg and Park (1991), and Evrard and Crone (1992).

In section 4 we consider a situation where the initial two point correlation is zero beyond a certain separation (i.e. $\xi(x) = 0$ for $x > x_0$). In the linear evolution the two point correlation will remain zero for $x > x_0$, but in the nonlinear evolution the two point correlation will develop because of the streaming of particles and the nonlocal nature of Newtonian gravity. This exhibits the influence of the small scales on the large scales. In section 5 we numerically investigate various cases where the initial two point function has a power law behaviour $x^{-\gamma}$ for large $x$. We look at three different quantities - the two point correlation, the average of the two point correlation over a sphere of radius $x$ and the pair velocity. We look at both the spatial and temporal behaviour of these quantities. In part I we found that at large separations the induced three point correlation function has the 'hierarchical form' for a class of initial conditions, but there are initial conditions where, because of the effect of the small scales on the large scales, the induced three point correlation does not have the hierarchical form. We investigate if the nonlinear correction to the two point correlation function shows any qualitative differences for these two kinds of initial conditions.

Hamilton et. al (1991) have suggested that in an $\Omega = 1$ universe the evolution of the two point correlation funtion can be described by a simple scaling relation. More recently Nityananda and Padmanabhan (1994) have examined the possible origin of this scaling relation. These arguments are based oninterpolation between the evolution in the linear regime and in the stable clustering regime. In section 5 we also check the validity of the proposed scaling relation in the weakly nonlinear regime.

In section 6 we discuss a possible interpretation for an interesting result we obtain in the numerical investigations of section 5. This interpretation is in terms of a simple diffusion process. In section 7 we present the conclusions.

## 2. Calculating the two point correlation.

### 2.1. Notation and the equation governing the two point correlation.

Throughout this paper we shall be using a comoving coordinate system. Instead of the cosmic time $t$, we shall use a parameter $\lambda$ defined by

$$d\lambda = \frac{dt}{S^2(t)} \tag{1}$$

where $S(t)$ is the cosmological scale factor. This parameter, which was also used in part I, was introduced by Doroshkevich et. al. (1980) and it make the calculations simpler. For an $\Omega = 1$ universe we have (part I)

$$S(\lambda) = \frac{3}{2\pi G\rho\lambda^2} \tag{2}$$

where $\rho$ is the mean comoving density. We shall use $\Omega = 1$ throughout this paper. The momentum of a particle $p_\mu$ is defined by

$$p_\mu = m\frac{d}{d\lambda}x_\mu \,. \tag{3}$$

We consider the initial density perturbation to be a Gaussian random field. This is completely described by the two point correlation function $\xi(x)$.

We also define a potential $\phi$ such that in the linear regime

$$\xi(x^1, x^2) = \frac{1}{2}\frac{\lambda_0^5}{\lambda^4}\nabla^4\phi\left(\mid x^1 - x^2 \mid\right) \sim \epsilon^2 \tag{4}$$

where $\lambda_0$ is the initial value of the parameter $\lambda$ and $\epsilon \sim (\delta\rho/\rho)$ is a small number in terms of which we are doing the perturbative expansion.

We define $f(p,\lambda), c(x^1,p^1,x^2,p^2,\lambda)$ and $d(1,2,3,\lambda)$ as the reduced ensemble averaged one, two and three point distribution functions in the appropriate phase spaces. Notice that we have used $(1,2,3)$ to indicate the point $(x^1,p^1,x^2,p^2,x^3,p^3)$ in phase space. Sometimes we shall use (1, 2, 3) to denote points in space. The difference will be clear from the context. The superscript indices a, b, c, etc. are used to indicate different positions.

We shall be using the various velocity moments of these distribution functions. Some of them are defined below and the notation for the rest follow the same pattern. We have

$$n = \int f(1)\, d^3p^1 \quad (nm = \rho)\,, \tag{5}$$

$$\int p^1_\mu f(1)\, d^3p^1 = 0\,, \tag{6}$$

where $n$ is the number density of particles,

$$n < \left(p^1_\mu\right)^2 >_1 \;=\; \int \left(p^1_\mu\right)^2 f(1)\, d^3p^1 \sim \epsilon^2\,, \tag{7}$$

$$n^2 \xi\left(x^1, x^2\right) \;=\; \int c(1,2)\, d^3p^1 d^3p^2 \sim \epsilon^2\,, \tag{8}$$

$$n^2 < p^a_\mu >_2 (x^1 x^2) \;=\; \int p^a_\mu c(1,2)\, d^3p^1 d^3p^2 \sim \epsilon^2 \tag{9}$$

$$n^2 < p^a_\mu p^b_\nu >_2 \left(x^1, x^2\right) \;=\; \int p^a_\mu p^b_\nu c(1,2)\, d^3p^1 d^3p^2 \sim \epsilon^2\,, \tag{10}$$

where a and b take values 1 and 2, and $\xi$ is the two point correlation function. The subscript outside the angular bracket indicates which distribution function the moment refers to e.g. $< p^a_\mu >_2 (1, 2, \lambda)$ is the first moment of the two point distribution function. The subscripts $(\mu, \nu, \alpha, etc)$ refer to Cartesian components and are to be summed over whenever they appear twice. Some more definitions are

$$n^3 \zeta\left(x^1, x^2, x^3\right) \;=\; \int d(1,2,3)\, d^3p^1 d^3p^2 d^3p^3 \sim \epsilon^3 \tag{11}$$

$$n^3 < p^a_\mu >_3 \left(x^1, x^2, x^3\right) \;=\; \int p^a_\mu d(1,2,3)\, d^3p^1 d^3p^2 d^3p^3 \sim \epsilon^3\,, \tag{12}$$

where $a$ takes values $1, 2$ and 3, and $\zeta(1, 2, 3)$ is the three point correlation function.

Also, we shall also use the following notation for the cartesian components of the inverse square force

$$X^{ab}_\mu = \frac{x^a_\mu - x^b_\mu}{\mid x^a - x^b \mid^3}. \tag{13}$$

We next present the equation governing the perturbative evolution of the two point correlation function. This equation, which was derived in part I, is

$$\frac{\partial^3}{\partial \lambda^3}\xi - 8\pi G\rho \left[S \frac{\partial}{\partial \lambda}\xi + \frac{\partial}{\partial \lambda}(S\xi)\right] = f_2 - f_3 - \frac{\partial}{\partial \lambda} f_1\,, \tag{14}$$

where,

$$f_1(1,2,\lambda) = SG\rho\frac{\partial}{\partial x^a_\mu}\int \zeta(1,2,3,\lambda)X^{3a}_\mu d^3x^3\,, \tag{15}$$

$$f_2(1,2,\lambda) = 2SGn\frac{\partial^2}{\partial x^b_\nu \partial x^a_\mu}\int < p^a_\mu >_3 (1,2,3,\lambda)X^{3b}_\nu d^3x^3\,, \tag{16}$$

$$f_3(1,2,\lambda) = \frac{1}{m^3}\frac{\partial^3}{\partial x^c_\sigma \partial x^b_\nu \partial x^a_\mu} < p^a_\mu p^b_\nu p^c_\sigma >_2 (1,2,\lambda)\,. \tag{17}$$

Here the position indices take values 1 and 2 and are to be summed when they appear twice.

For an $\Omega = 1$ universe this becomes

$$\frac{\partial^3}{\partial\lambda^3}\xi - \frac{24}{\lambda^2}\frac{\partial}{\partial\lambda}\xi + \frac{24}{\lambda^3}\xi = f_2 - f_3 - \frac{\partial}{\partial\lambda}f_1\,,\,. \tag{18}$$

We separately consider the various terms on the right hand side of equation (18). We first consider equation (15). This depends on three point correlation function $\zeta$ which has been calculated to order $\epsilon^4$ in part I.

We reproduce the expression below. We have

$$\begin{aligned}\zeta(1,2,3,\lambda) \;&=\; \frac{1}{28}\left(\frac{\lambda_0}{\lambda}\right)^8\lambda_0^2\sum_{a=1}^{3}\left[3\frac{\partial}{\partial x^a_\mu}\left(\nabla^4\phi\left(a,a_1^{'}\right)\frac{\partial}{\partial x^a_\mu}\nabla^2\phi\left(a,a_2^{'}\right)\right)\right.\\ &+\left. 2\frac{\partial^2}{\partial x^a_\mu \partial x^a_\nu}\left(\frac{\partial}{\partial x^a_\nu}\left(\nabla^2\phi\left(a,a_1^{'}\right)\right)\frac{\partial}{\partial x^a_\mu}\left(\nabla^2\phi\left(a,a_2^{'}\right)\right)\right)\right]\,.. \end{aligned} \tag{19}$$

In the equation for the three point correlation function the following conventions are used
A. the position indices, e.g. $a$, take values 1,2, and 3 corresponding to the corners of the triangle for which the three point correlation function is being evaluated. Also, a position index which appears twice or more should be summed over the allowed values.
B. for a fixed value of the position index (e.g. $a = 1$), $a_1^{'}$ and $a_2^{'}$ are to be summed over the other two values (i.e. $a_1^{'} = 2, a_2^{'} = 3$ and $a_1^{'} = 3, a_2^{'} = 2$). This is to be done whenever such a combination of three position indices appear.

In some of the equations for the other moments of the three point distribution function, if indicated, the summation convention A may not hold, but the convention B always holds.

To calculate $f_2$ and $f_3$ we have to first calculate the following quantities: $< p^a_\mu >_3 (1,2,3,\lambda)$ and $\partial^a_\mu\partial^b_\nu\partial^c_\sigma < p^a_\mu p^a_\nu p^a_\sigma >_2 (1,2,\lambda)$.. These calculations are are discussed in the following two subsections and the final result is presented in subsection 2.4. The overall strategy is the same as in part I. We take velocity moments of the BBGKY hierarchy and retain terms only up to order $\epsilon^4$. As a result of this we obtain a set of ordinary differential equations in the parameter $\lambda$. These equations have complicated spatial dependences. The spatial calculation is simplified by taking spatial derivatives (curl and divergence) and we then obtain a set of equations that can be easily solved simultaneously.

### 2.2. The triplet momentum.

The triplet momentum $< p^a_\mu >_3 (1,2,3,\lambda)$ is defined as the first moment of the three-point distribution function $d\,(1,2,3,\lambda)$.

$$\int p^a_\mu d\,(1,2,3,\lambda)\, d^3p^1 d^3p^2 d^3p^3 = n^3 < p^i_\mu >_3 \left(x^1, x^2, x^3, \lambda\right) . \tag{20}$$

It is a function of three positions 1,2 and 3, and the index '$a$' in $< p^a_\mu >_3$, which indicates at which vertex of the triangle we are considering the momentum, can refer to any one of them.

We want to calculate this quantity to order $\epsilon^4$. This is the lowest order for which it has non-zero value for Gaussian initial conditions.

The evolution of the triplet momentum is governed by the first moment of the third equation of the BBGKY hierarchy

$$\begin{aligned}
&\frac{\partial}{\partial\lambda} < p^a_\mu >_3 + \frac{1}{m}\frac{\partial}{\partial x^b_\nu} < p^a_\mu p^b_\nu >_3 (1,2,3,\lambda) \\
- \;& SmG\rho \int \zeta\left(a'', 4\right) X^{4a}_\mu d^3x^4 - SmG\rho \int \chi\,(1,2,3,4) X^{4a}_\mu d^3x^4 \\
- \;& SmG\rho\xi\left(a, a_1'\right) \int \xi\left(a_2', 4\right) X^{4a}_\mu d^3x^4 = 0\,.
\end{aligned} \tag{21}$$

To evaluate $< p^a_\mu >_3$ we separately consider both its curl and divergence with respect to $x^a$ and use these to construct it. All the equations given below for the curl and divergence are valid only to order $\epsilon^4$ and in all of them the summation convention A does not hold. In all these equations the indices $a, b$ and $c$ refer to the three different corners of the triangle that we are considering i.e. $a \neq b \neq c$.

The curl of equation (21) is

$$\begin{aligned}
&\frac{\partial}{\partial\lambda} F^a_\beta + \frac{1}{m}\sum_{b\neq a} G^{ab}_\beta \\
+ \;& \frac{5m}{2}\left(\frac{\lambda_0}{\lambda}\right)^{10} \epsilon_{\beta\mu\nu}\left[\partial^a_\mu \nabla^4\phi\left(a, a_1'\right) \partial^a_\nu \nabla^2\phi\left(a, a_2'\right)\right] = 0
\end{aligned} \tag{22}$$

where,

$$F^a_\beta(1,2,3,\lambda) = \epsilon_{\beta\mu\nu}\partial^a_\mu < p^a_\nu >_3 \tag{23}$$

and

$$G^{ab}_\beta(1,2,3,\lambda) = \epsilon_{\beta\mu\nu}\partial^a_\mu\partial^b_\sigma < p^a_\nu p^b_\sigma >_3 . \tag{24}$$

and we have used the fact that for Gaussian initial conditions to order $\epsilon^4$

$$< p^a_\mu p^a_\nu >_3 = < p^a_\mu >_2 \left(a, a_1'\right) < p^a_\mu >_2 \left(a, a_2'\right) \tag{25}$$

to evaluate $G^{aa}_{\beta}$.

The divergence of equation (21) is

$$\frac{\partial}{\partial\lambda}J^a + \frac{1}{m}\sum_{b\neq a}K^{ab} + m\left(\frac{\lambda_0}{\lambda}\right)^{10}\left[\sum_{q=1}^{3}\frac{3}{14}g\left(q\right) + \frac{1}{2}g\left(a\right)\right] = 0 \tag{26}$$

where

$$J^a\left(1,2,3,\lambda\right) = \partial^a_\mu < p^a_\mu >_3 , \tag{27}$$

$$K^{ab}\left(1,2,3,\lambda\right) = \partial^a_\mu \partial^b_\nu < p^a_\mu p^b_\nu >_3 \tag{28}$$

and,

$$\begin{aligned} g(a) \;=\; & 5\nabla^4\phi\left(a,a_1^{'}\right)\nabla^4\phi\left(a,a_2^{'}\right) + 7\partial^a_\mu\nabla^2\phi\left(a,a_1^{'}\right)\partial^a_\mu\nabla^4\phi\left(a,a_2^{'}\right) \\ & +2\partial^a_\mu\partial^a_\nu\nabla^2\phi\left(a,a_1^{'}\right)\partial^a_\mu\partial^a_\nu\nabla^2\phi\left(a,a_2^{'}\right) . \end{aligned} \tag{29}$$

We have used equation (26) to evaluate $K^{aa}$ and equation (19) for the three point correlation function.

Next we consider the second moment of the three point distribution function

$$\begin{aligned} & \frac{\partial}{\partial\lambda} < p^a_\mu p^b_\nu >_3 + \frac{1}{m}\frac{\partial}{\partial x^c_\sigma} < p^a_\mu p^b_\nu p^c_\sigma >_3 \left(1,2,3,\lambda\right) \\ - \; & \frac{Sm^2G}{n^3}\int\left(\delta^{ca}_{\sigma\mu}p^b_\nu + \delta^{cb}_{\sigma\nu}p^a_\mu\right) f\left(c\right) d\left(c^{''},3\right) X x^{4c}_\sigma d^3x^4 d^{12}p \\ - \; & \frac{Sm^2G}{n^3}\int\left(\delta^{ca}_{\sigma\mu}p^b_\nu + \delta^{cb}_{\sigma\nu}p^a_\mu\right) c\left(c,c_1^{'}\right) c\left(c_2^{'},4\right) X^{4c}_\sigma d^3x^4 d^{12}p \\ - \; & \frac{Sm^2G}{n^3}\int\left(\delta^{ca}_{\sigma\mu}p^b_\nu + \delta^{cb}_{\sigma\nu}p^a_\mu\right) e\left(1,2,3,4\right) X^{4c}_\sigma d^3x^4 d^{12}p = 0 . \end{aligned} \tag{30}$$

We use this to get an equation for $G^{ab}_\beta$

$$\begin{aligned} \frac{\partial}{\partial\lambda}G^{ab}_\beta \; & + \; \frac{1}{m}H^{abc}_\beta + \frac{6m}{\lambda^2}F^a_\beta \\ & + \; 5m^2\frac{\lambda_0^{10}}{\lambda^{11}}\epsilon_{\beta\mu\nu}\left[\partial^a\nu\nabla^2\phi\left(a,b\right)\partial^a_\mu\nabla^4\phi\left(a,c\right)\right. \\ & + \; \left.\partial^a_\mu\nabla^4\phi\left(a,b\right)\partial^a_\nu\nabla^2\phi\left(a,c\right)\right] = 0 . \end{aligned} \tag{31}$$

where

$$H^{abc}_\beta = \epsilon_{\beta\mu\nu}\partial^a_\mu\partial^b_\alpha\partial^c_\sigma < p^a_\nu p^b_\alpha p^c_\sigma >_3 \tag{32}$$

We have used the fact that for Gaussian initial conditions to order $\epsilon^4$

$$\begin{aligned} < p^a_\mu p^a_\nu p^b_\sigma >_3 \; & = \; < p^a_\mu p^b_\sigma >_2 \left(a,b\right) < p^a_\nu >_2 \left(a,c\right) \\ & + \; < p^a_\nu p^b_\sigma >_2 \left(a,b\right) < p^a_\mu >_2 \left(a,c\right) \end{aligned} \tag{33}$$

to evaluate $H_\beta^{abd}$ when $d = a$ or $d = b$.

By taking divergences of equation (30) we obtain

$$\frac{\partial}{\partial\lambda}K^{ab} + \frac{1}{m}L^{abc} + \frac{6m}{\lambda^2}\left(J^a + J^b\right) + m^2\frac{\lambda_0^{10}}{\lambda^{11}}\left[g\left(a\right) + g\left(b\right)\right] = 0 \tag{34}$$

where

$$L^{abc} = \partial_\mu^a \partial_\nu^b \partial_\sigma^c < p_\mu^a p_\nu^b p_\sigma^c >_3 \ . \tag{35}$$

Finally we have the third moment of the equation for the three point distribution function

$$\begin{aligned}
&\frac{\partial}{\partial\lambda} < p_\mu^a p_\nu^b p_\sigma^c >_3 + \frac{1}{m}\frac{\partial}{\partial x_\gamma^d} < p_\mu^a p_\nu^b p_\sigma^c p_\gamma^d >_3 (1,2,3,\lambda) \qquad (36)\\
- &\frac{Sm^2G}{n^3}\int\left(\delta_{\gamma\mu}^{ea}p_\nu^b p_\sigma^c + \delta_{\gamma\nu}^{eb}p_\mu^a p_\sigma^c + \delta_{\gamma\sigma}^{ec}p_\mu^a p_\nu^b\right) f\left(e\right) d\left(e^{"},4\right) X_\gamma^{4e} d^3x^4 d^{12}p\\
- &\frac{Sm^2G}{n^3}\int\left(\delta_{\gamma\mu}^{ea}p_\nu^b p_\sigma^c + \delta_{\gamma\nu}^{eb}p_\mu^a p_\sigma^c + \delta_{\gamma\sigma}^{ec}p_\mu^a p_\nu^b\right) c\left(e,e_1^{'}\right) c\left(e_2^{'},4\right) X_\gamma^{4e} d^3x^4 d^{12}p\\
- &\frac{Sm^2G}{n^3}\int\left(\delta_{\gamma\mu}^{ea}p_\nu^b p_\sigma^c + \delta_{\gamma\nu}^{eb}p_\mu^a p_\sigma^c + \delta_{\gamma\sigma}^{ec}p_\mu^a p_\nu^b\right) e\left(1,2,3,4\right) X_\gamma^{4e} d^3x^4 d^{12}p = 0\,.
\end{aligned}$$

This can be used to obtain the equation for $H_\beta^{abc}$

$$\begin{aligned}
\frac{\partial}{\partial\lambda}H_\beta^{abc} &+ \frac{6m}{\lambda^2}\left(G_\beta^{ab} + G_\beta^{ac}\right)\\
&+ 10m^3\left(\frac{\lambda_0^{10}}{\lambda^{12}}\right)\epsilon_{\beta\mu\nu}\left[\partial_\nu^a\nabla^2\phi\left(a,b\right)\partial_\mu^a\nabla^4\phi\left(a,c\right)\right.\\
&+ \left.\partial_\mu^a\nabla^4\phi\left(a,b\right)\partial_\nu^a\nabla^2\phi\left(a,c\right)\right] = 0
\end{aligned} \tag{37}$$

To obtain this equation we have used the fact that for Gaussian initial conditions to order $\epsilon^4$

$$< p_\mu^a p_\nu^a p_\sigma^b p_\gamma^c >_3 = < p_\mu^a p_\sigma^b >_2 < p_\nu^a p_\gamma^c >_3 + \text{permutations}\,. \tag{38}$$

Taking divergence of equation (37) we obtain

$$\frac{\partial}{\partial\lambda}L^{abc} + \frac{6m}{\lambda^2}\left(K^{ab} + K^{bc} + K^{ca}\right) + 2\sum_{p=1}^{3} m^3\left(\frac{\lambda_0^{10}}{\lambda^{12}}\right) g\left(p\right) = 0\,. \tag{39}$$

The equations (22),(31) and (37) can be simultaneously solved to obtain

$$F_\beta^a = \frac{m}{2}\left(\frac{\lambda_0^{10}}{\lambda^9}\right)\epsilon_{\beta\mu\nu}\left[\partial_\mu^a\nabla^4\phi\left(a,a_1^{'}\right)\partial_\nu^a\nabla^2\phi\left(a,a_2^{'}\right)\right] \tag{40}$$

$$G_\beta^{ab} = m^2\left(\frac{\lambda_0^{10}}{\lambda^{10}}\right)\epsilon_{\beta\mu\nu}\left[\partial_\mu^a\nabla^4\phi\left(a,a_1^{'}\right)\partial_\nu^a\nabla^2\phi\left(a,a_2^{'}\right)\right] \tag{41}$$

$$H_\beta^{abc} = 2m^3\left(\frac{\lambda_0^{10}}{\lambda^{11}}\right)\epsilon_{\beta\mu\nu}\left[\partial_\mu^a\nabla^4\phi\left(a,a_1^{'}\right)\partial_\nu^a\nabla^2\phi\left(a,a_2^{'}\right)\right]\ . \tag{42}$$

Simultaneously solving equations (26),(34) and (39) we have

$$J^a = \frac{m}{14}\left(\frac{\lambda_0^{10}}{\lambda^9}\right)[2g(a)+g(b)+g(c)] \tag{43}$$

$$K^{ab} = \frac{m^2}{7}\left(\frac{\lambda_0^{10}}{\lambda^{10}}\right)[2g(a)+2g(b)+g(c)] \tag{44}$$

$$L^{abc} = \frac{4m^3}{7}\left(\frac{\lambda_0^{10}}{\lambda^{11}}\right)[g(a)+g(b)+g(c)]\,. \tag{45}$$

Using these we obtain the triplet momentum as

$$\begin{aligned}
&< p_\mu^a >_3 (1,2,3,\lambda) = m\frac{\lambda_0^{10}}{\lambda^9}\left[\frac{1}{2}\partial_\mu^a\nabla^2\phi\left(a,a_1^{'}\right)\nabla^4\phi\left(a,a_2^{'}\right)\right.\\
+&\ \frac{1}{7}\partial_\mu^a\left(\partial_\nu^a\nabla^2\phi\left(a,a_1^{'}\right)\partial_\nu^a\nabla^2\phi\left(a,a_2^{'}\right)\right)\\
+&\ \frac{5}{7}\partial_\mu^a\nabla^2\phi\left(a,a_1^{'}\right)\nabla^4\phi\left(a_1^{'},a_2^{'}\right)+\frac{1}{2}\partial_\mu^a\partial_\nu^{a_1^{'}}\nabla^2\phi\left(a,a_1^{'}\right)\partial_\nu^{a_1^{'}}\nabla^2\phi\left(a_1^{'},a_2^{'}\right)\\
+&\ \frac{1}{2}\partial_\mu^a\partial_\nu^{a_1^{'}}\phi\left(a,a_1^{'}\right)\partial_\nu^{a_1^{'}}\nabla^4\phi\left(a_1^{'},a_2^{'}\right)+\frac{2}{7}\partial_\mu^a\partial_\nu^{a_1^{'}}\partial_\sigma^{a_1^{'}}\phi\left(a,a_1^{'}\right)\partial_\nu^{a_1^{'}}\partial_\sigma^{a_1^{'}}\nabla^2\phi\left(a_1^{'},a_2^{'}\right)\\
-&\ \left.\frac{3}{56\pi}\int X_\mu^{4a}\partial_\nu^4\left(\partial_\nu^4\nabla^2\phi\left(4,a_1^{'}\right)\nabla^4\phi\left(4,a_2^{'}\right)\right)d^3x^4\right]
\end{aligned} \tag{46}$$

and we also obtain for $a \neq b$

$$\begin{aligned}
&\partial_\nu^b < p_\mu^a p_\nu^b >_3 (1,2,3,\lambda) = \frac{m^2}{7}\frac{\lambda_0^{10}}{\lambda^{10}}\left[7\partial_\mu^a\nabla^2\phi\left(a,a_1^{'}\right)\nabla^4\phi\left(a,a_2^{'}\right)\right.\\
+&\ 2\partial_\mu^a\left(\partial_\nu^a\nabla^2\phi\left(a,a_1^{'}\right)\partial_\nu^a\nabla^2\phi\left(a,a_2^{'}\right)\right)+20\partial_\mu^a\nabla^2\phi(a,b)\nabla^4\phi(b,c)\\
+&\ 10\partial_\mu^a\nabla^2\phi(a,c)\nabla^4\phi(b,c)+14\partial_\mu^a\partial_\nu^b\nabla^2\phi(a,b)\partial_\nu^b\nabla^2\phi(b,c)\\
+&\ 14\partial_\mu^a\partial_\nu^b\phi(a,b)\partial_\nu^b\nabla^4\phi(b,c)+7\partial_\mu^a\partial_\nu^c\nabla^2\phi(a,c)\partial_\nu^c\nabla^2\phi(b,c)\\
+&\ 7\partial_\mu^a\partial_\nu^c\phi(a,c)\partial_\nu^c\nabla^4\phi(b,c)+8\partial_\mu^a\partial_\nu^b\partial_\sigma^b\phi(a,b)\partial_\nu^b\partial_\sigma^b\phi(b,c)\\
+&\ 4\partial_\mu^a\partial_\nu^c\partial_\sigma^c\phi(a,c)\partial_\nu^c\partial_\sigma^c\phi(b,c)\\
-&\ \left.\frac{3}{4\pi}\int X_\mu^{4a}\partial_\nu^4\left(\partial_\nu^4\nabla^2\phi\left(4,a_1^{'}\right)\nabla^4\phi\left(4,a_2^{'}\right)\right)d^3x^4\right]
\end{aligned} \tag{47}$$

### 2.3. The third moment of the two point distribution function.

In this section the position indices takes the values 1 and 2, and they are to be summed whenever they appear twice or more.

The third moment of the two point distribution function is governed by the equation

$$\frac{\partial}{\partial\lambda} < p_\mu^a p_\nu^b p_\sigma^c >_2 + \frac{1}{m}\frac{\partial}{\partial x_\gamma^d} < p_\mu^a p_\nu^b p_\sigma^c p_\gamma^d >_2 (1,2,\lambda)$$

$$- \frac{Sm^2G}{n^2} \int \left( \delta^{ea}_{\gamma\mu} p^b_\nu p^c_\sigma + \delta^{eb}_{\gamma\nu} p^a_\mu p^c_\sigma + \delta^{ec}_{\gamma\sigma} p^a_\mu p^b_\nu \right) f(e)\, c\left(e', 3\right) X^{3e}_\gamma d^3x^3 d^9p$$

$$- \frac{Sm^2G}{n^2} \int \left( \delta^{ea}_{\gamma\mu} p^b_\nu p^c_\sigma + \delta^{eb}_{\gamma\nu} p^a_\mu p^c_\sigma + \delta^{ec}_{\gamma\sigma} p^a_\mu p^b_\nu \right) d(1,2,3) X^{3e}_\gamma d^3x^3 d^9p = 0\,.$$

If we take divergence with respect to all the three sets of free indices we have

$$\begin{aligned} & \frac{\partial}{\partial\lambda} \partial^a_\mu \partial^b_\nu \partial^c_\sigma < p^a_\mu p^b_\nu p^c_\sigma >_2 + \frac{1}{m} \partial^a_\mu \partial^b_\nu \partial^c_\sigma \partial^d_\gamma < p^a_\mu p^b_\nu p^c_\sigma p^d_\gamma >_2 \\ + \; & 12\pi SmG\rho \left[ < p^a_\mu p^a_\nu >_1 \partial^a_\mu \partial^a_\nu \xi(1,2,\lambda) + \partial^a_\mu \partial^a_\nu < p^a_\mu p^a_\nu >_2 (1,2,\lambda) \right] \\ - \; & 3SmG\rho \partial^a_\mu \partial^b_\nu \partial^c_\sigma \int X^{3c}_\sigma < p^a_\mu p^b_\nu >_3 (1,2,3,\lambda) d^3x^3 = 0\,. \end{aligned} \tag{48}$$

For Gaussian initial conditions the terms in this equation have non-zero values only at order $\epsilon^4$ and higher. To order $\epsilon^4$ there are two unknown functions in the equation i.e. $\partial^a_\mu \partial^b_\nu \partial^c_\sigma < p^a_\mu p^b_\nu p^c_\sigma >_2$ and $< p^a_\mu p^a_\nu >_2$. We need one more equation to self-consistently solve this equations. This equation is obtained from the second moment of the two point distribution function

$$\begin{aligned} \frac{\partial}{\partial\lambda} & < p^a_\mu p^b_\nu >_2 (1,2,\lambda) + \frac{\partial}{m\,\partial x^c_\sigma} < p^a_\mu p^b_\nu p^c_\sigma >_2 (1,2,\lambda) \\ & - \frac{m^2GS}{n^2} \int \left( \delta^{ca}_{\sigma\mu} p^b_\nu + \delta^{cb}_{\sigma\nu} p^a_\mu \right) f(c)\, c\left(c', 3\right) X^{3c}_\sigma d^3x^3 d^9p \\ & - \frac{m^2GS}{n^2} \int \left( \delta^{ca}_{\sigma\mu} p^b_\nu + \delta^{cb}_{\sigma\nu} p^a_\mu \right) d(1,2,3) X^{3c}_\sigma d^3x^3 d^9p = 0\,. \end{aligned} \tag{49}$$

From equation (49) we get

$$\begin{aligned} & \frac{\partial}{\partial\lambda} \partial^a_\mu \partial^a_\nu < p^a_\mu p^a_\nu >_2 + \frac{1}{3m} \partial^a_\mu \partial^b_\nu \partial^c_\sigma < p^a_\mu p^b_\nu p^c_\sigma >_2 \\ + \; & \frac{2}{3m} \partial^a_\mu \partial^a_\nu \partial^a_\sigma < p^a_\mu p^a_\nu p^a_\sigma >_2 - 2SmG\rho \partial^a_\mu \partial^a_\nu \int < p^a_\mu >_3 X^{a3}_\nu d^3x^3 = 0\,. \end{aligned} \tag{50}$$

Simultaneously solving equations (48) and (50) we obtain

$$\begin{aligned} & \partial^a_\mu \partial^b_\nu \partial^c_\sigma < p^a_\mu p^b_\nu p^c_\sigma >_2 = \frac{3}{26} \partial^a_\mu \partial^a_\nu \partial^a_\sigma < p^a_\mu p^a_\nu p^a_\sigma >_2 + \frac{5\lambda}{52m} \partial^a_\mu \partial^b_\nu \partial^c_\sigma \partial^d_\gamma < p^a_\mu p^b_\nu p^c_\sigma p^d_\gamma >_2 \\ + \; & \frac{45m}{26\lambda} < p^a_\mu p^a_\nu >_1 \partial^a_\mu \partial^a_\nu \xi(1,2,\lambda) - \frac{27m^2}{52\lambda^2\pi} \partial^a_\mu \partial^a_\nu \int < p^a_\mu >_3 X^{3a}_\nu d^3x^3 \\ - \; & \frac{45m}{104\lambda\pi} \partial^a_\mu \partial^b_\nu \partial^c_\sigma \int < p^a_\mu p^b_\nu >_3 X^{3c}_\sigma d^3x^3 \end{aligned} \tag{51}$$

All the terms on the left hand side of equation (51) are known to order $\epsilon^4$. Writing it in terms of $\phi$ we have

$$\partial^a_\mu \partial^b_\nu \partial^c_\sigma < p^a_\mu p^b_\nu p^c_\sigma >_2 = m^3 \left( \frac{\lambda_0^{10}}{\lambda^{11}} \right) \left[ -\frac{82}{7} \nabla^6 \phi(1,2) \nabla^2 \phi(0) \right.$$

$$
\begin{aligned}
&+ \quad \frac{240}{7}\nabla^4\phi(1,2)\nabla^4\phi(1,2) + \frac{816}{7}\partial^1_\mu\nabla^4\phi(1,2)\partial^1_\mu\nabla^2\phi(1,2) \\
&+ \quad 96\partial^1_\mu\partial^1_\nu\nabla^2\phi(1,2)\partial^1_\mu\partial^1_\nu\nabla^2\phi(1,2) + \frac{528}{7}\partial^1_\mu\partial^1_\nu\partial^1_\alpha\nabla^2\phi(1,2)\partial^1_\mu\partial^1_\nu\partial^1_\alpha\phi(1,2) \\
&+ \quad 48\partial^1_\mu\partial^1_\nu\nabla^4\phi(1,2)\partial^1_\mu\partial^1_\nu\phi(1,2) + \frac{96}{7}\partial^1_\mu\partial^1_\nu\partial^1_\alpha\partial^1_\beta\phi(1,2)\partial^1_\mu\partial^1_\nu\partial^1_\alpha\partial^1_\beta\phi(1,2) \\
&- \quad \frac{1}{7\pi}\partial^1_\mu\partial^1_\nu\int X^{31}_\mu\Big(36\partial^1_\nu\nabla^2\phi(1,3)\nabla^4\phi(2,3) + \frac{45}{2}\partial^1_\nu\partial^3_\alpha\nabla^2\phi(1,3)\partial^3_\alpha\nabla^2\phi(2,3) \\
&+ \quad \frac{45}{2}\partial^1_\nu\partial^3_\alpha\phi(1,3)\partial^3_\alpha\nabla^4\phi(2,3) + 9\partial^1_\nu\partial^3_\alpha\partial^3_\beta\phi(1,3)\ \partial^3_\alpha\partial^3_\beta\nabla^2\phi(2,3)\Big)\,d^3x^3\Big] \qquad (52)
\end{aligned}
$$

### 2.4. The two point correlation.

Using the results derived in the two previous sections we can calculate $f_2$ and $f_3$. We first consider only the $\lambda$ dependence of $\xi$. Equation (18) may be written as

$$
\frac{\partial^3}{\partial\lambda^3}\xi - \frac{24}{\lambda^2}\frac{\partial}{\partial\lambda}\xi + \frac{24}{\lambda^3}\xi = \left(\frac{\lambda_0}{\lambda}\right)^{11}\left[f_2(\lambda_0) - f_3(\lambda_0) + \frac{10}{\lambda_0}f_1(\lambda_0)\right], \qquad (53)
$$

This has a solution

$$
\xi(1,2,\lambda) = -\frac{1}{504}\frac{\lambda_0^{11}}{\lambda^8}\left[f_2(\lambda_0) - f_3(\lambda_0) + \frac{10}{\lambda_0}f_1(\lambda_0)\right]. \qquad (54)
$$

Using equations (15),(16) and (17) we calculate the spatial dependence of the right hand side. This gives us

$$
\begin{aligned}
\xi^{(2)}(1,2,\lambda) &= \frac{1}{504}\left(\frac{\lambda_0^{10}}{\lambda^8}\right)\Big[-32\nabla^6\phi(1,2)\nabla^2\phi(0) - 14\nabla^4\phi(1,2)\nabla^4\phi(0) \\
&+ \quad \frac{900}{7}\nabla^4\phi(1,2)\nabla^4\phi(1,2) + 360\partial^1_\mu\nabla^4\phi(1,2)\partial^1_\mu\nabla^2\phi(1,2) \\
&+ \quad \frac{1602}{7}\partial^1_\mu\partial^1_\nu\nabla^2\phi(1,2)\partial^1_\mu\partial^1_\nu\nabla^2\phi(1,2) + 144\partial^1_\mu\partial^1_\nu\partial^1_\alpha\nabla^2\phi(1,2)\partial^1_\mu\partial^1_\nu\partial^1_\alpha\phi(1,2) \\
&+ \quad 126\partial^1_\mu\partial^1_\nu\nabla^4\phi(1,2)\partial^1_\mu\partial^1_\nu\phi(1,2) + \frac{144}{7}\partial^1_\mu\partial^1_\nu\partial^1_\alpha\partial^1_\beta\phi(1,2)\partial^1_\mu\partial^1_\nu\partial^1_\alpha\partial^1_\beta\phi(1,2) \\
&- \quad \frac{1}{\pi}\int X^{31}_\mu\Big(\frac{3}{2}\partial^1_\nu\nabla^2\phi(1,3)\partial^2_\mu\partial^2_\nu\nabla^4\phi(2,3) + \frac{3}{2}\partial^1_\alpha\partial^1_\nu\nabla^2\phi(1,3)\partial^2_\mu\partial^2_\alpha\partial^2_\nu\nabla^2\phi(2,3) \\
&+ \quad \frac{15}{2}\partial^1_\alpha\partial^1_\nu\phi(1,3)\partial^2_\mu\partial^2_\alpha\partial^2_\nu\nabla^4\phi(2,3) + 3\partial^1_\alpha\partial^1_\beta\partial^1_\nu\phi(1,3)\partial^2_\mu\partial^2_\alpha\partial^2_\beta\partial^2_\nu\nabla^2\phi(2,3) \\
&- \quad 15\nabla^4\phi(1,3)\partial^2_\mu\nabla^4\phi(2,3) - \frac{21}{2}\partial^1_\nu\nabla^4\phi(1,3)\partial^2_\nu\partial^2_\mu\nabla^2\phi(2,3)\Big)\,d^3x^3\Big] \qquad (55)
\end{aligned}
$$

This is the two point correlation function at order $\epsilon^4$.

We do a Fourier transform of (55) and compare it with the result obtained in the single stream approximation (Makino et. al 1992) and find that the two match.

The algebra involved in deriving equation (55) was checked using the mathematical package MATHEMATICA and the Fourier transform was done using this package.

The calculation presented here, which is based on the equations of the BBGKY hierarchy, has the effects of multistreaming, if any, at the lowest order of non-linearity i.e. $\epsilon^4$. This matches with the results obtained in the single stream approximation which does not take into account any effect of multistreaming. Hence we find that there are no effects of multistreaming at this order of non-linearity.

Given the initial two point correlation function $\xi^{(1)}(x,\lambda)$, to evaluate the non-linear correction at order $\epsilon^4$ we solve the equation

$$\nabla^2\left(\nabla^2\phi(x)\right) = \frac{2}{\lambda_0}\xi^1(x,\lambda_0) \tag{56}$$

to obtain $\nabla^2\phi(x)$ and then solve for $\phi(x)$ and use these in equation (55). These calculations are simplified a lot if we use the fact that $\xi(x,\lambda_0)$ is spherically symmetric. Equation (56) does not uniquely determine the functions $\nabla^2\phi(x)$ and $\phi(x)$. If $\nabla^2\phi(x)$ is a solution of equation (56) then so is $\nabla^2\phi(x)+C$ where $C$ is some constant. Under this transformation we also have $\phi(x)\to\phi(x)+\frac{Cx^2}{6}$. If we consider all the terms in equation (55) that are affected by this

$$\begin{aligned} &- \quad 32\nabla^6\phi(1,2)\,\nabla^2\phi(0) + 126\partial^1_\mu\partial^1_\nu\nabla^4\phi(1,2)\,\partial^1_\mu\partial^1_\nu\phi(1,2) \\ &- \quad \frac{15}{2\pi}\int X^{31}_\mu\partial^1_\nu\partial^1_\alpha\phi(1,3)\,\partial^2_\mu\partial^2_\nu\partial^2_\alpha\nabla^4\phi(2,3)\,d^3x^3 \end{aligned} \tag{57}$$

we find that the correction to the two point correlation function is unchanged by such transformations and independent of $C$. Similarly, we can add a constant to $\phi(x)$ but this obviously does not have any affect as only derivatives of $\phi(x)$ appear in equation (55). Hereafter we shall use the boundary condition that $\nabla^2\phi(x)$ should vanish as $x$ goes to infinity to fix the constant $C$.

At this stage we should point out that it such a choice of $C$ is not always convenient. For example if the initial two point correlation is such that the power spectrum has the form $P(k)\propto k^n$ with $n\le -1$ at small $k$, then the boundary condition stated above implies that

$$\nabla^2\phi(0) = -4\pi\int_0^\infty P(k)dk \tag{58}$$

is infinite. Although $\nabla^2\phi(0)$ is infinite and this quantity appears in the non-linear correction to the two point correlation function $\xi^{(2)}(x,t)$, we may still get a finite $\xi^{(2)}(x,t)$ for certain initial conditions. This is because now $\xi^{(2)}$ is the difference of two infinite large quantities which cancel out to give a finite result. The same problem is encountered if one does the analysis in Fourier space where for $-1\le n\ge -3$ the correction to the power spectrum is a finite quantity which is the difference of two divergent integrals (Vishniac 1983). In real space this situation is easily handled by changing the boundary condition used to calculate $\nabla^2\phi(x)$. If we use the boundary condition $\nabla^2\phi(0)=0$ to fix the constant $C$ than the situation discussed above does not occur and it is possible to calculate $\xi^{(2)}(x,t)$ solely in terms of finite quantities. Here we shall only deal with situations where the former boundary condition ( $\lim_{x\to\infty}\nabla^2\phi(x)=0$) can be applied.

## 3. The pair velocity

We next calculate the first moment of the two point distribution function, $< p^a >_2 (1,2)$, to order $\epsilon^4$ . This is a function of two positions and the index $a$ indicates at which of the two positions the momentum is being considered. We use the pair continuity equation

$$\frac{\partial}{\partial\lambda}\xi\,(1,2,\lambda) + \frac{1}{m}\frac{\partial}{\partial x^a_\mu} < p^a_\mu >_2 (1,2,\lambda) = 0\,. \tag{59}$$

to obtain

$$< p^a_\mu >_2 (1,2,\lambda) = \frac{m}{8\pi}\int \partial^a_\mu\left(\frac{1}{\mid y - y^{'}\mid}\right)\frac{\partial}{\partial\lambda}\xi(y^{'})d^3y^{'}\,, \tag{60}$$

where

$$y^a_\mu = x^a_\mu - x^{a^{'}}_\mu$$

end $a^{'}$ refers to the complement of $a$ (e.g. if $a$=1, $a^{'}$ =2). Using the two point correlation at order $\epsilon^4$ calculated in the previous section we get at order $\epsilon^4$

$$\begin{aligned}
< p^a_\mu >_2 \;=\;& \frac{m}{126}\frac{\lambda_0^{10}}{\lambda^9}\left[\frac{y^a_\mu}{y^3}\int_0^y \Big[-32\nabla^6\phi(x)\nabla^2\phi\,(0) - 14\nabla^4\phi(x)\nabla^4\phi\,(0)\right.\\
+\;& \frac{900}{7}\nabla^4\phi(x)\nabla^4\phi(x) + 360\partial_\mu\nabla^4\phi(x)\partial_\mu\nabla^2\phi(x)\\
+\;& \frac{1602}{7}\partial_\mu\partial_\nu\nabla^2\phi(x)\partial_\mu\partial_\nu\nabla^2\phi(x) + 144\partial_\mu\partial_\nu\partial_\alpha\nabla^2\phi(x)\partial_\mu\partial_\nu\partial_\alpha\phi(x)\\
+\;& 126\partial_\mu\partial_\nu\nabla^4\phi(x)\partial_\mu\partial_\nu\phi(x) + \frac{144}{7}\partial_\mu\partial_\nu\partial_\alpha\partial_\beta\phi(x)\partial_\mu\partial_\nu\partial_\alpha\partial_\beta\phi(x)\Big]\,x^2dx\\
-\;& \frac{3}{\pi}\int X^{3a^{'}}_\alpha\Big[\frac{1}{2}\partial^{a^{'}}_\beta\nabla^2\phi\left(a^{'},3\right)\partial^a_\alpha\partial^a_\beta\partial^a_\mu\nabla^2\phi\,(a,3)\\
+\;& \frac{1}{2}\partial^{a^{'}}_\beta\partial^{a^{'}}_\gamma\nabla^2\phi\left(a^{'},3\right)\partial^a_\alpha\partial^a_\beta\partial^a_\gamma\partial^a_\mu\phi\,(a,3)\\
+\;& \frac{5}{2}\partial^{a^{'}}_\beta\partial^{a^{'}}_\gamma\phi\left(a^{'},3\right)\partial^a_\alpha\partial^a_\beta\partial^a_\gamma\partial^a_\mu\nabla^2\phi\,(a,3)\\
-\;& \partial^{a^{'}}_\alpha\partial^{a^{'}}_\beta\partial^{a^{'}}_\gamma\partial^{a^{'}}_\delta\phi\left(a^{'},3\right)\partial^a_\beta\partial^a_\gamma\partial^a_\delta\partial^a_\mu\phi\,(a,3)\\
-\;& 5\nabla^4\phi\left(a^{'},3\right)\partial^a_\alpha\partial^a_\mu\nabla^2\phi\,(a,3)\\
-\;& \left.\frac{7}{2}\partial^{a^{'}}_\beta\nabla^4\phi\left(a^{'},3\right)\partial^a_\alpha\partial^a_\beta\partial^a_\mu\phi\,(a,3)\Big]\,d^3x^3\right]\,.
\end{aligned} \tag{61}$$

A quantity related to the first moment of the two point distribution function is the pair current

$$j_\mu(x,\lambda) = \frac{< p^2_\mu >_2 - < p^1_\mu >_2}{m} \tag{62}$$

whose divergence gives the rate at which the correlation at any separation is growing. We use this to calculate the pair velocity which is the ensemble average of the relative peculiar velocity between any two particles at a comoving separation $x$ at time $t$ (or $\lambda$).

In terms of the pair current this is

$$v_\mu(x,t) = S\frac{d\lambda}{dt}\left[\frac{j_\mu(x,\lambda)}{1+\xi(x,\lambda)}\right] \tag{63}$$

## 4. Effect of small scales on large scales.

In this section we study how the small scales affect the large scales. We consider a situation where initially $\xi^1(x) = 0$ for $x > x_0$. In the linear regime $\xi(x)$ will continue to have value zero for $x > x_0$. This will not be true in the nonlinear regime and there will be nonzero correlation for $x > x_0$ due to the streaming of particles and the non-local nature of gravity. In this section we want to find out, to the lowesst order of nonlinearity, the nature of the induced correlations.

We first consider an extreme case of the above situation where the initial two point correlation is a Dirac delta function and $\xi^1(x) = 0$ for $x > 0$. We initially have

$$\xi^{(1)}(x,\lambda_0) = \frac{1}{2}\lambda_0\nabla^4\phi(x) = \frac{1}{2}\lambda_0 A\delta^3(x). \tag{64}$$

Solving for the potential we get

$$\phi(x) = -\frac{A}{8\pi}x \tag{65}$$

and using this in equation (55) we obtain for $x > 0$.

$$\xi^{(2)}(x,\lambda) = \frac{5A^2}{98\pi^2}\frac{\lambda_0^{10}}{\lambda^8}\frac{1}{x^6}\,. \tag{66}$$

This shows the influence of the small scales on the large scales.

We next write equation (66) in a form that can be compared with a general case where the initial two point correlation function has compact support. This is done by introducing a quantity $M(\lambda)$ which is related to the integral of $\xi^{(1)}(x,\lambda)$ over the whole region $x \le x_0$ where it is non-zero i.e.

$$M(\lambda) = \int \xi^{(1)}(x,\lambda)x^2dx\,. \tag{67}$$

We can then write equation (66) for $x \gg x_0$ as

$$\xi^{(2)}(x,\lambda) = 3.264\frac{M^2(\lambda)}{x^6}\,. \tag{68}$$

The general case where the initial two point correlation has compact support cannot be treated analytically because the integrals in equation (55) cannot be done analytically.

To check whether the conclusions drawn from the case where the initial condition is a delta function can be generalized we have considered a situation where

$$\nabla^4 \phi(x) = 16 - 12x + x^3 \quad \text{for} \quad \mathrm{x} \leq 2 \tag{69}$$

end

$$\nabla^4 \phi(x) = 0 \quad \text{for} \quad \mathrm{x} > 2\,. \tag{70}$$

This corresponds to the self-convolution of a spheres of unit radius. The corresponding power spectrum has the form

$$P(k) \propto \left[\frac{\sin(k) - k\ \cos(k)}{k^3}\right]^2 \tag{71}$$

Doing the integrals in equation ( 55) numerically we find that for large $x$ *i.e.* in the interval $30 \leq x \leq 40$, the induced two point correlation can be fitted by the form

$$\xi^{(2)}(x, \lambda) = 3.277 \frac{M^2(\lambda)}{x^6}\,. \tag{72}$$

For larger values of $x$ the integrals could not be reliably evaluated numerically as the numbers involved are extremely small. The initial and induced two point correlation are shown in figure 1.

We propose that the form (68) holds for the two point correlation induced ar large distances in all cases where the initial two point correlation has compact support and a nonzero total integral.

We next consider the pair current. The initial pair current for the delta function initial condition is

$$j_\mu(x, \lambda) = \frac{4}{\lambda} \frac{x_\mu}{x^3} M(\lambda) \tag{73}$$

This current is divergence free everywhere except at the origin. Since it does not change the correlation at any nonzero separation it may be thought of as matter flowing into the overdense regions from infinity and flowing out of the overdense regions to infinity (Zeldovich and Novikov 1983, section 10.4).

If we try to calculate the nonlinear correction to the pair current for the delta function initial condition we get a divergent answer. This is because the integral $\int_0^\infty \xi^{(2)}(y) y^2 dy$ diverges .

For the other initial condition considered above (equation 71) we find numerically that the induced pair current (figure 2) has a $x^{-2}$ behaviour at large $x$. This is the leading part of the induced pair current and it too represents the flow of matter from infinity. In addition to this divergence-free part, the pair current also has a part that has a $x^{-5}$ behaviour at large $x$ and it corresponds to a local redistribution of matter. It is the latter that gives rise to the $x^{-6}$ correlation at large $x$.

For large separations ($x > 2$) the pair velocity has the same spatial dependence as the pair current.

Peebles (1980) and Zeldovich and Novikov (1983) (and references therein) have discussed the effect of small scales on large scales. They have considered an initial power spectrum which has the value zero at small $k$ (large length scales). They find that the rearrangement of matter on the small scales gives rise to a power spectrum of the form $P(k) \propto k^4$ at small $k$ or large scales. N-body simulations by Mellot and Shandarin (1990) find that there is indeed a $k^4$ power spectrum generated in the initial stages of the evolution but this changes over to a $k^3$ form by the time the small scales saturate. The initial condition considered in these calculations corresponds to a situation where the universe can be divided into small cells and the matter content in all the cells is exactly equal. Also the matter in each cell is distributed such that its center of mass coincides with the center of the cell and the total momentun of each cell is zero. They have investigated the effect of redistributing the matter inside each cell, conserving the mass and momentum of the individual cells. Matter and momentum conservation are of course also built into the scheme used in this paper, but the analysis presented here differs in that the initial conditions used here are such that the cells all have different masses and the mass fluctuations of cells that are at large separations are uncorrelated. These are not the same initial conditions that give rise to the $k^4$ spectrum. We have studied the nature of the correlations that develops between the fluctuation in the mass of cells that are at a large separation. Note that this is a quantity in real space whereas the $k^4$ spectrum being analytic in the limit $k \to 0$ does not imply any long range power law correlation in real space (e.g. $P(k) = k^4 e^{-k^2}$ has a Fourier transorm $\nabla^4 e^{-x^2/4}$. Hence the results given here are not directly related to those of Peebles (1980).

## 5. Numerical Investigations.

The expressions for the contribution to the two point correlation and the pair velocity at order $\epsilon^4$ are rather complicated. To understand them better we proceed to evaluate them numerically for different initial conditions.

### 5.1. The spatial behaviour of the $\epsilon^4$ contributions.

First we consider initial two point correlations such that the corresponding power spectrum has the form $P(k) = k^n e^{-k}$. We consider cases where n takes the values $n = .5, 1, 1.5, 2$ and 3. These initial conditions have just one length scale which is introduced by the exponential cut-off which becomes effective for $k > 1$. In all these cases the correlation function has a power law behaviour $x^{-\gamma}$ for large separations and the corresponding values of the index $\gamma$ are $\gamma = 3.5, 4, 4.5, 6$ and 6. We also consider the case $P(k) = ke^{-k^2}$ which has a Gaussian cut-off for the power spectrum at large $k$ instead of the exponential cut-off used in all the other cases.

Figures 3 shows the function $\xi^{(1)}(x)$ for two of the cases. This function is defined so that at any instant the two point correlationn at order $\epsilon^2$ is

$$\xi^{(1)}(x,t) = \left(\frac{S(t)}{S(t_0)}\right)^2 \xi^{(1)}(x). \tag{74}$$

Figure 3 also shows the quantity $\overline{\xi^{(1)}}(x)$. This when multiplied by $\left(\frac{S(t)}{S(t_0)}\right)^2$ gives the $\sim \epsilon^2$ contribution to $\bar{\xi}(x,t)$ which is defined as

$$\bar{\xi}(x,t) = \frac{3}{x^3}\int_0^x \xi(y,t) y^2 dy. \tag{75}$$

This is the average of the two point correlation function over a sphere of radius $x$ . All of the cases considered here satisfy the relation $\xi^{(1)}(x) \propto \overline{\xi^{(1)}}(x)$ for large $x$. This is crucial in deciding the behaviour of the induced three point correlation function $\zeta(1,2,3)$ at large separations and for this class of initial conditions we have $\zeta(1,2,3) \sim \xi(1,2)\xi(2,3)$ + permutations.

Figures 4 shows the function $v^{(1)}(x)$ which is related to the radial component of the pair velocity at order $\epsilon^2$ at any instant by the relation

$$v^{(1)}(x,t) = S(t)\frac{d}{dt}\left(\frac{S(t)}{S(t_0)}\right)^2 v^{(1)}(x)\,. \tag{76}$$

For all these cases the initial pair velocity too has a power law behaviour $x^{-\beta}$ at large separations and we have $\beta = \gamma - 1$ as expected.

It should be pointed out that the initial conditions when $n = 3$ differs from the other cases. For all the other cases $\xi^{(1)}(x)$ crosses zero only once, it is positive for small $x$ and goes over to a negative value at lerge $x$ where it has the power law form. The quantities $\overline{\xi^{(1)}}(x)$ and $v^{(1)}(x)$ do not change sign , the former is positive and the latter is negative. In the case when $n = 3$, $\xi^{(1)}(x)$ crosses zero twice, and $\overline{\xi^{(1)}}(x)$ and $v^{(1)}(x)$ cross zero once. Thus in this case ($n = 3$) at large $x$ the signs of all the quantities are opposite to the signs in the other cases.

We have calculated the $\epsilon^4$ contribution to all of the above mentioned quantities in the renge $0 \le x \le 40$.

We first discuss the large separation behaviour of $\xi^{(2)}(x)$ which is defined such that the $\epsilon^4$ contribution to the two point correlation is $\left(\frac{S(t)}{S(t_0)}\right)^4 \xi^{(2)}(x)$. This function is shown in figure 5. For all the cases we find that at large separations $\xi^{(2)}(x)$ has a power law form $\xi^{(2)}(x) \sim x^{-\eta}$ with $\eta = \gamma - 2$ i.e. $\xi^{(2)}(x) \sim \nabla^2\xi^{(1)}(x)$. Motivated by this we investigated whether there is any simple relation between $\xi^{(2)}(x)$ and $\xi^{(1)}(x)$ which holds for all the cases. We looked at the ratio

$$Z = \frac{\xi^{(2)}(x)}{\nabla^2\phi(0)\nabla^6\phi(x)} \tag{77}$$

at large $x$ for the different cases and we find that the value of $Z$ is nearly the same $(-.048 < Z < -.049)$ for all the cases. This relation can also be expressed as

$$\xi^{(2)}(x) = R\left[\int_0^\infty \xi^{(1)}(y) y dy\right] \nabla^2\xi^{(1)}(x) \tag{78}$$

with $R = -4Z \approx .194$. In terms of the power spectrum this may be written as

$$P_2(k) = -\frac{R}{2\pi^2}\left[\int_0^\infty P_1(k^{'})dk^{'}\right]k^2 P_1(k) \tag{79}$$

where $P_1(k)$ is the initial power spectrum and $P_2(k)$ is the correction to the power spectrum at $\epsilon^4$. The fact that $\int_0^\infty P(k)dk > 0$ tells us that $P_2(k) < 0$. In real space we can say that $\xi^{(2)}(x)$ has the same sign as $\xi^{(1)}(x)$.

Based on this numerical evidence we make a hypothesis that equation (78) holds for all initial conditions where the initial power spectrum index $n$ satisfies $0 < n$. We have not considered any cases where $n$ is negative. For such cases ($n \leq -1$) the integral in equation (79) does not converge and we do not expect this relation to hold.

One of the factors we believe to be responsible for the spread in the values of $R$ (or $Z$) is that $\xi^{(2)}(x)$ goes to the power law form asymptotically and for all the cases we have not been able to calculate $\xi^{(2)}(x)$ to equally large separations. As of now we have no rigorous derivation for equation (78), but we give a heuristic interpretation in terms of a diffusion process in section 6.

Makino et. al. (1992) have analytically calculated $P_2(k)$ for various power law initial conditions, of which the case where

$$P(k)_1 = A\frac{k}{k_c} \quad \text{for} \quad k \leq k_c \tag{80}$$

and zero elsewhere is of interest to us. In the limit $k- > 0$ their analytic expression for $P_2(k)$ reduces to

$$P_2(k) = -\frac{61}{315\,(2\pi)^2}A^2 k^3\,. \tag{81}$$

This can be written as

$$P_2(k) = -\frac{61}{315}\left(\frac{1}{2\pi^2}\right)\left[\int_0^\infty P_1(k^{'})dk^{'}\right]k^2 P_1(k) \tag{82}$$

which on comparison with equation (79) gives us $R = \frac{61}{315}$ or $Z = -\frac{1}{4}\left(\frac{61}{315}\right) = -.0484$.

We see that this matches equation (79) and this serves as a test of our hypothesis.

At small $x$ the behaviour of $\xi^{(2)}(x)$, as shown in figure 5 is rather complicated and is difficult to generalize. We find that it starts with a positive value at $x = 0$, which falls fast, becomes negative and oscillates around zero a few times before going over to the power law fowm.

We next consider the behaviour of $\bar{\xi}(x,t)$. At large separations this has a behaviour quite similar to $\xi^{(2)}(x)$ as shown in figures 5 and we find that

$$\overline{\xi^{(2)}}(x) = R\left[\int_0^\infty \xi^{(1)}(y)y dy\right]\frac{1}{x}\nabla^2(x\overline{\xi^{(1)}}(x))\,. \tag{83}$$

We see that at large separations $\overline{\xi^{(2)}}(x)$ has the same sign as $\overline{\xi^{(1)}}(x)$. The small $x$ behaviour of $\overline{\xi^{(2)}}(x)$ is similar to $\xi^{(2)}(x)$.

The behaviour of $v^{(2)}(x)$, which is defined such that the $\epsilon^4$ contribution to the pair velocity is $S(t)\frac{d}{dt}S(t)^4 v^{(2)}(x)$, is shown in figure 6. For large values of $x$ the behaviour can be described by

$$v^{(2)}(x) = R\left[\int_0^\infty \xi^{(1)}(y)y dy\right]\nabla^2 v^{(1)}(x)\ . \tag{84}$$

We see that $v^{(2)}(x)$ too has the same sign as $v^{(1)}(x)$ at large $x$. At small $x$ the function $v^{(2)}(x)$ starts from zero and rises fast and then falls off and changes sign a few times before going over to the power law form.

We see that at very small scales the $\epsilon^4$ contribution acts to increase the correlation, whereas at intermediate scales it can act to increase oe decrease the correlations..

We next consider a case where the initial power spectrum has the form $P_1(k) = e^{-k}$. The two point correlation function has the form $\xi^{(1)}(x) \sim x^{-4}$ for large $x$ but $\overline{\xi^{(1)}}(x)$ does not have the same behaviour and we have $\overline{\xi^{(1)}}(x) \sim x^{-3}$ instead (figure 7). Because of this, at large separations the three point correlation function exhibits a behaviour which is quite different from the one exhibited by the cases considered previously and in this case the three point correlation, as discussed in part I, does not have the 'hierarchical form'. We find that the large $x$ behaviour of $\xi^{(2)}(x)$ too is somewhat different as compared to the previous cases. Although we find that $\xi^{(2)}(x) \propto \nabla^2\xi^{(1)}(x)$ the factor relating the two is diferent from that found for the previous cases and if we fit a formula like equation (79) we get $R \approx .496$ instead of .194.

We next consider the large $x$ behaviour of $\overline{\xi^{(2)}}(x)$. We see that it has a behaviour of the form $x^{-6}$ as compared with the initial function $\overline{\xi^{(1)}}(x)$ which has the form $x^{-3}$ i.e. a difference of three in the power law index. This is different from the previous cases where there was a difference of two between the index for the linear function and the nonlinear correction as seen in equation (83). This can be easily understood by noting that if we try to relate $\overline{\xi^{(2)}}(x)$ with $\overline{\xi^{(1)}}(x)$ using an expression like equation (83) we find that the fact that $\overline{\xi^{(1)}}(x) \sim x^{-3}$ implies that the right hand side is zero. We then deduce that, in addition to a part that behaves as $x^{-3}$, $\overline{\xi^{(1)}}(x)$ has part that behaves as $x^{-4}$. At large $x$ the value of $\overline{\xi^{(1)}}(x)$ is determined solely by the former term as the latter fall off much faster, but the behaviour of the nonlinear correction $\overline{\xi^{(2)}}(x)$ is determined by the latter as the first term does not contribute. We find that equation (83) gives a good fit for $R \approx .496$ which is consistent with the fit for $\xi^{(2)}(x)$.

At large $x$ the behaviour of $v^{(2)}(x)$ is similar to the behaviour of $\overline{\xi^{(2)}}(x)$.

At small $x$ both $\xi^{(2)}(x)$ and $\overline{\xi^{(2)}}(x)$ start of with positive values whichh fall off fast. At intermediate values $\xi^{(2)}(x)$ changes sign twice and goes over to the the power law form at large $x$ whereas $\overline{\xi^{(2)}}(x)$ changes sign only once and hence the two quantities have opposite signs at large $x$. The behaviour of $v^{(2)}(x)$ is similar to the behaviour of $\xi^{(2)}(x)$ axcept that it starts from the value zero.

### 5.2. The temporal behaviour

Here we would like to investigate the evolution of the correlation function and the pair velocity. It is generally believed that the linear results should hold on some legthscale until the density contrast averaged over that length scale is of the order of unity. Having calculated the lowest order nonlinear term we can see when this becomes of the order of the linear term. This would be a different criterion for determining when the linear results would no longer be applicable. Here we wish to compare these two criteria and investigate whether they are the same.

Before proceeding further we should remind the reader that we are working in the continuum (or fluid ) limit and the initial conditions are such that the perturbative treatment is valid at all length scales. As a result of this there is a growth of clustering even at the smallest scales. Although in reality there are bound particles at some extremely small scales these are not taken into account.

We only discuss the case with $P_1(k) \propto ke^{-k}$ here. We find that the other cases considered have a similar behaviour.

The smaller scales go nonlinear first. We first consider the evolution of $\xi(0,t)$ which is the mean square density fluctuation (figure 8). We find that the nonlinear correction enhances the growth of the correlation. We also find that the linear term is equal to the correction when $\xi(0,t)$ is of the order of unity. This is as expected and here it happens at $S(t) = .121$. At $S(t) = .053$ the correction is one tenth of the linear term and we may expect that the other higher order corrections not considered here will not contribute before this epoch.

Next we consider the separation $x = .1$. Figure 9 shows the evolution of the various quantities of interest. There is no qualitative difference here with $x = 0$ except that at $x = 0$ the pair velocity is zero. When $x = 1$ (figure 10), there is a qualitative difference. Previously the effect of the nonlinear terms was to increase the correlation and now it tends to decrease it, but here too $\overline{\xi^{(1)}}(x,t)$ and $\overline{\xi^{(2)}}(x,t)$ are equal when they both are of order unity. At very large separations e.g. $x = 20$ (figure 11) the behaviour is quite different. We find that $\overline{\xi^{(1)}}(x,t)$ and $\overline{\xi^{(2)}}(x,t)$ are equal when $\overline{\xi}(x,t) \sim .01$. The correlation function and the pair velocity show a similar behaviour too. Hence it appears that at large scales the linear theory is breaking down much before one would expect it to. It should be noted that this happens at a very large value of $S(t)$ ($\sim 10$) and the perturbative approach has broken down much earlier on the small scales. The correction $\overline{\xi^{(2)}}(x,t) = R\left[\int_o^\infty \xi^{(1)}(y,t)y dy\right] \nabla^2 \xi^{(1)}(x,t)$ is nonlocal and it has a relatively large contribution from the small scales. In this expression the contribution from the small scale keeps on growing, whereas in reality the clustering saturates at the small scales because of virialization. Peebles (1980) has argued that once virialized objects have formed on some small scales, those small scales have no influence on the evolution of the large scales. Taking this into account we see that at large scales the lowest order nonlinear correction overestimates the contribution from the small scales in the epoch when the small scales have become highly nonlinear and have formed structures which are in virial equilibrium. We are still left with the fact that any particular small scale affects the evolution of the large scales until the small scale in question gets virialized. Just how significant this effect is has to be estimated by some independent means (i.e. not lowest order nonlinear perturbation). One possible method of doing this is to compare the perturbative results with N-body simulations. Little, Weinberg and Park (1991), and Evrard and Crone (1992) have used N-body simulations to study how various scales affect one another in the non-linear epoch for an initial power spectrum of the form $P(k) \propto k^{-1}$. They find that the large scales have a strong influence on the evolution of the small scales but the small scales have very little influence on the evolution of the large scales. This is not in direct conflict with the findings of this paper because they are not claimed to be valid for the $n = -1$ type spectra.

In a more recent paper Jain and Bertschinger (1994) have studied how the various scales influence one another in the nonlinear epoch for the CDM initial conditions where the linear power spectra has $P_1(k) \propto k$ for small $k$. They have calculated the lowest order nonlinear correction to the power spectrum $P_2(k)$ and they compare it to the results from N-body simulations. They follow the evolution until the density contrast on the scale of $8h^{-1}Mpc$ is of the order of unity and they find that at small $k$ the nonlinear correction to the power spectrum is negative and is much smaller than the linear power spectrum $(i.e. P_2(k,t) \ll P_1(k,t)$, which is in good agreement with their N-body simulations. We have considered the CDM spectrum with the normalisation used by Jain and Bertschinger (1994) and used it in equation (79) to calculate the lowest order nonlinear correction to the power spectrum, The results, not given in detail here, are consisitent with those obtained by Jain and Bertschinger(1994) $(i.e. P_2(k,t) \ll P_1(k,t)$ . For the standard CDM initial conditions with the normalization $\sigma_8 = 1$, the condition $\overline{\xi^{(1)}}(x,t) \sim \overline{\xi^{(2)}}(x,t) \ll 1$ at large $x$ will occur only if we extend the evolution into the future.

In section 6 we discuss a possible physical picture for the influence of the small scales on the large scales. We also discuss a possible means of independently estimating how significant this effect will be for general initial conditions.

### 5.3. Scaling relations.

Hamilton et. al. (1991) have suggested that in an $\Omega = 1$ universe the evolution of the two point correlation funtion can be described by a simple universal relation whose exact form they have obtained by fitting N-body simulations. More recently Nityananda and Padmanabhan (1994) have examined the possible origin of this universal scaling relation. The scaling relation can be based on the conjecture that the dimensionless pair velocity $h(a,x) = -\frac{v(x,t)}{\dot{S}(t)x}$ depends on $S(t)$ and $x$ through $\bar{\xi}(x,t)$ alone. This conjecture is valid in the linear regime. Here we perturbatively test this conjecture at the lowest order of nonlinearity.

We look at the behaviour of $h(x,t)$ and $\bar{\xi}(x,t)$ in some range where we are reasonably sure that the perturbative approach gives a good description of the clustering. Since the smallest scales go nonlinear first, the criterion is based on the properties at $x = 0$ and we restrict our analysis to the epoch when $\xi^{(2)}(o,t) \leq .1\xi^{(2)}(0,t)$. Because of this conservative criterion we find that $h(x,t)$ differs very little from the linear value of $\frac{2}{3}\bar{\xi}(x,t)$. To look at the nature of this small deviation we consider the ratio $\frac{h(x,t)}{\bar{\xi}(x,t)}$. Figure 12 shows this ratio as a function of $\bar{\xi}(x,t)$ for different $S(t)$ for the initial condition $P_1(k) \propto ke^{-k}$. We find that for a fixed value of $\bar{\xi}(x,t)$ the ratio has a spread of values. Although this spread is not large (5 percent), it is comparable to the largest deviation of mean square density from the linear prediction (ten percent). We also find that this ratio has a systematic behaviour which can be understood by looking at the same quantity as a function of $x$ for different $S(t)$ (figure 13). We see that there are points where the corrections to the pair velocity $v^{(2)}(x)$ and also to $\overline{\xi^{(2)}}(x)$ are both zero and $h(x,t)$ and $\bar{\xi}(x,t)$ there (figure 12) continues to follow the linear evolution. If we consider the first such point, then for smaller values of $x$ (larger values of $\bar{\xi}(x,t)$) the deviation is positive with respect to the linear value. For larger values of $x$ (smaller values of $\bar{\xi}(x,t)$) the deviation is negative with respect to the linear value. At large x (beyond the second zero crossing of the correction to the pair velocity) the deviation again becomes positive, but this is not seen in figure 12 because the whole range of $x$ gets mapped to a very small range of $\bar{\xi}(x,t)$ between 0 and some very small value. Based on this we draw the conclusion that we cannot express $h(x,t)$ as a function of $\bar{\xi}(x,t)$ alone.

We have carried out this excercise for all the different initial conditions discusssed earlier and they too exhibit a simiiar behaviour.

As was discussed earlier, equations (83) and (84) give a good description of $\overline{\xi^{(2)}}(x)$ and $v^{(2)}(x)$ at large $x$ for a large set of initial conditions of the form $\xi^{(1)}(x) \propto x^{-\gamma}$. Using these and the fact that $\frac{v^{(1)}(x)}{x} = -\frac{\overline{\xi^{(1)}}(x)}{3}$ we have

$$h(x,t) = 2S(t)^2 \left[1 + \frac{2S(t)^2\gamma(\gamma-1)R}{x^2} \int_0^\infty \xi^{(1)}(y)ydy\right] \frac{x^{-\gamma}}{3} \tag{85}$$

and

$$\bar{\xi}(x,t) = S(t)^2 \left[1 + \frac{S(t)^2\gamma(\gamma-1)R}{x^2} \int_0^\infty \xi^{(1)}(y)ydy\right] x^{-\gamma} \; . \tag{86}$$

This also show that $h(x,t)$ is not a function of $\bar{\xi}(x,t)$ alone.

## 6. An interpretation based on diffusion.

In the previous section we saw that for a certain class of initial conditions equation (78) gives a good fit for $\xi^{(2)}(x)$ in terms of $\xi^{(1)}(x)$. In this section we provide a possible interpretation for this equation. The cosmic energy equation (Peebles 1980 and reference therein), which is the second moment of the first equation of the BBGKY hierarchy, allows us to relate the integral that appears in equation (78) to the mean square momentum. In the notation used in this paper the cosmic energy equation can be written as

$$\frac{\partial}{\partial\lambda} < (p^1)^2 >_1 (\lambda) = 4\pi SG\rho m^2 \frac{\partial}{\partial\lambda} \int_o^\infty \xi(x,\lambda)xdx \tag{87}$$

where $x =\mid x^1 - x^2 \mid$.

Instead of looking at the evolution of $< (p^1)^2 >$, we consider the motion of the particles as a function of the growing mode which in this case is the scale factor $S(t)$. We define a velocity

$$u_\mu = \frac{dx_\mu}{dS} = \frac{1}{m}\frac{d\lambda}{dS}p_\mu \tag{88}$$

and $< u^2 > (S(t))$ is the mean square of this velocity. In terms of this the cosmic energy equation is

$$\frac{\partial}{\partial S}\left[S^3 < u^2 >\right] = \frac{3S}{2}\frac{\partial}{\partial S}\int_o^\infty \xi(x,S)xdx \tag{89}$$

which in the linear regime gives us

$$< u^2 >= \frac{1}{S(t_0)^2}\int_o^\infty \xi^{(1)}(x)xdx, \tag{90}$$

i.e. $< u^2 >$ does not change. Note that in this equation we have not explicitly shown the superscript (1) over $u$ to indicate that it the linear part of $u$. Henceforth we shall use $u$ for the linear part of the same quantity. Using equation (90) we can write equation (78) as

$$\xi^{(2)}(x,t) = RS^2(t) < u^2 > \left(\frac{S(t)}{S(t_0)}\right)^2 \nabla^2\xi^{(1)}(x) \tag{91}$$

which looks like the solution of a diffusion equation. It is in light of this that we interpret equation (78).

Consider a particular realization of the Gasussian density fluctuation field. We consider two points $x^1$ and $x^2$ where the density fluctuations are $\Delta(x^1,t)$ and $\Delta(x^2,t)$ respectively. The $\Delta(x,t)$ of the fluid element at any point $x$ grows according to linear theory and the fluid element moves according to some random velocity $u_\mu$ which is assumed uncorrelated to the velocity at any other point. We then have

$$\begin{aligned}\Delta(x,t) &= \left(\frac{S(t)}{S(t_0)}\right)^2 \Delta(x+Su,t_0) = \left(\frac{S(t)}{S(t_0)}\right)^2 [\Delta(x,t_0) \\ &+ S(t)u_\mu\partial_\mu\Delta(x,t_0) + \frac{S^2(t)}{2}u_\mu u_\nu\partial_\mu\partial_\nu\Delta(x,t_0)\Big] .\end{aligned} \tag{92}$$

Using this and taking an ensemble average we have

$$\begin{aligned}\xi(x^1-x^2,t) &= < \Delta(x^1,t)\Delta(x^2,t) >= \left(\frac{S(t)}{S(t_0)}\right)^2 \xi(x^1-x^2,t_0) \\ &+ \frac{1}{3}S^2(t) < u^2 > \left(\frac{S(t)}{S(t_0)}\right)^2 \nabla^2\xi(x^1-x^2,t_0)\end{aligned} \tag{93}$$

where

$$< u_\mu u_\nu >= \frac{\delta_{\mu\nu}}{3} < u^2 > . \tag{94}$$

The first term on the right hand side corresponds to the linear growth and the second term is the lowest nonlinear correction. We see that by this process we have a nonlinear correction which matches with equation (91) except for the numerical factor $R$. This difference can be attributed to the fact that we have kept the velocity $u_\mu$ constant and this is only valid in the linear regime. In the nonlinear evolution this is not true. Based on equation (93) we suggest that the simple diffusion process described above gives a description of the evolution of the two point correlation at the lowest order of nonlinearity for large separations. Thus, when we are looking at the correlation at a large separation $x$, we can consider the initial perturbation field to be growing linearly, but there is one more effect to be taken into account. This effect is the local rearrangement of matter on small scales and this may be thought of as being random. The local rearrangement is on a lengthscale $L \sim S(t)\sqrt{< u^2 >}$ and we expect this picture to be valid when $L \ll x$. When the mean displacement becomes comparable to the separation ($i.e. L \sim x$) we can longer treat the displacements as random and this picture is not valid.

We see that $L^2(t)$, the mean square displacement of the particles from their initial positions is the quantity that determones the influence of the small scales on the large scales. It is also this quantity that gets overestimated in the later epochs in the perturbative result. To make an independent estimate of the influence of the small scales on the large scales it is necessary to estimate this quantity by some other means (possibly N-body simulations).

## 7. Conclusions.

We find that the lowest order nonlinear ($\sim \epsilon^4$) correction to the two point correlation calculated using the BBGKY hierarchy matches with the result obtained using the single stream approximation. Thus at this order there are no effects of multistreaming. The calculations presented here do not tell us if we we can get any effect due to multistreaming by going to higher orders of perturbation or whether the inability to follow the transition from a single streamed flow to a multistreamed flow is a limitation of the perturbative approach. In a separate study (Bharadwaj 1995) we have considered s system whose dynamics is governed be the Zel'dovich approximation instead of the gravitational dynamics. The transition from a single streamed flow to a multi-streamed flow occurs in the Zel'dovich approximation also. For this system we consider situations such that the initial flow is single streamed and it can be described by the gradient of a potential which is a Gaussian random field. We use two methods to perturbatively follow the evolution of the two point correlation function 1. The single streamed fluid equations which break down once multi-streaming occurs, and 2. distribution functions in phase space which are in principle valid even in the multi-streamed epoch. We find that the results obtained using these two methods match at all orders of perturbation indicating that there are no effects of multi-streaming at any order in a perturbative expansion. Non-perturbative analysis shows that the effects of multi-streaming is of the form $\frac{1}{S(t)}e^{-\frac{A}{S^2(t)}}$ and hence they do not contribute in a perturbative expansion. We expect such conclusions to hold for the gravitational dynamics also.

It is known that if the initial power spectrum at small $k$ has the form $k^n$ with $-3 < n < -1$, then the lowest order nonlinear correction to the power spectrum is a finite quantity which is the difference of two infinite quantities. This is inconvenient if one wants to numerically evaluate the correction to the power spectrum. We find that in real space, i.e. for the two point correlation function, this difficulty can be avoided by changeing the boundary conditions for the Laplacian operator. Usually when solving an equation of the form $\nabla^2 f(x) = g(x)$ one uses the boundary condition $\mathrm{limit}_{x->\infty}\mathrm{f(x)} = 0$. We find that if we use the boundary condition $f(0) = 0$ then the above situation does not arise.

To study the influence of small scales on the large scales we have considered two cases where the two point correlation is initially zero at large separations(i.e. $\xi^{(1)}(x) = 0$ for $x > x_0$). One of the cases has been treated analytically and the other numerically. We find that at large separations the induced two point correlation function has the form $\xi^{(2)}(x) \propto x^{-6}$. We also find that in both the cases the constant of proportionality is the same functional of the initial two point correlation function.

We have numerically investigated some cases where the initial two point correlation has a power law form $x^{-\gamma}$ at large $x$. The cases we have studied have a power spectrum of the form $k^n$ with $0 \leq n \leq 3$ at small $k$ and have an exponential or gaussian cut off at large $k$. The cut off introduces a length scale and we find that at scales much smaller than this scale the nonlinear term enhances the growth of clustering. At intermediate scales we find that the nonlinear term changes sign more than once and it can act both ways i.e. to increase or decrease the clustering.

At large scales we find that the behaviour of the nonlinear term depends on the condition whether $\overline{\xi^{(1)}}(x) \propto \xi^{(1)}(x)$ or not. For the cases where this condition is satisfied we find that equation (78) gives a good fit to the lowest order nonlinear correction $\xi^{(2)}(x)$ to the two point correlation. We find similar equations for the average of the two point correlation $\overline{\xi^{(2)}}(x)$ and the pair velocity $v^{(2)}(x)$ too. We have interpreted equation (78) in terms of a simple diffusion process. For all the quantities the nonlinear term has the same sign as the corresponding linear term.

For the case where $\overline{\xi^{(1)}}(x)$ is not proportional to $\xi^{(1)}(x)$ we find that we obtain an equation similar to equation (78) with a different numerical coefficient. The quanitities $\overline{\xi^{(2)}}(x)$ and $v^{(2)}(x)$ also have a similar behaviour, but for both of them it is not the leading linear part that contributes.

In all the cases equation (78) shows the effect of the various scales on the large scales. Equation (78) has an integral of the initial two point correlation over all scales, and for the initial conditions that we have considered the small scales contribute the most. Thus we see that at the lowest order of nonlinearity the small scales effect the large scales.

Hansel et al (1985) have studied the effect of large scales on the small scales and they find that in the weakly nonlinear regime the small scale perturbations get modulated by the large scale perturbations. This effect is like a diffusion in Fourier space because the effect of the disturbances at small $k$ is to spread out the power spectrum at large $k$. It is very interesting that the effect of small scales on the large scales is a diffusion process in real space and the effect of the large scales on the small scales is like a diffusion process in Fourier space.

It is believed that perturbation theory is valid at a certain scale until $\bar{\xi}(x,t)$ at that scale is of the order of one. Another criterion for the applicability of perturbation theory is that the second order term should be smaller than the linear term. We have compared these two criteria and we find that the two break down at nearly the same epoch at small scales. At large scales we find that the second order term becomes of the same magnitude as the linear term at an epoch when $\bar{\xi}(x,t) \ll 1$. This happens because the second order term at large scales is influenced by the small scales. This happens at an epoch which is much later than the epoch when the perturbative treatment breaks down at small scales and because of the coupling of scales the second order term may be an overestimate of the actual nonlinear effects at large scales. For the CDM spectrum with the standard normalization this will occur only if we continue the evolution into the future. We also propose that for general initial conditions the true mean square displacement of the particles from their original positions rather than their linear value may be used to obtain a more realistic estimate of the influence of small scales on the large scales.

We have tested the hypothesis that the dimensionless pair velocity $h(x,t)$ is an universal function of the average of the two point correlation $\bar{\xi}(x,t)$. In the linear regime we have $h(x,t) = \frac{2}{3}\bar{\xi}(x,t)$. We find that the lowest order nonlinear correction introduces length scale dependent deviations from this relation and the pair velocity cannot be expressed as a universal function of the average of the two point correlation function. Although our investigation is valid only in the weakly nonlinear regime where the departures from the scaling relation are small, it indicates that as the evolution proceeds the contribution from the lowest order nonlinear terms will tend to enhance this deviation from the scaling relation. The approximate scaling relations found in N-body simulations by Hamilton et. al. (1991) seems to indicate that as the evolution proceeds, effects that have not been considered in our analysis tend to cancel out the deviations predicted by the lowest oreder nonlinear terms. This could be due to either higher order nonlinear effects or non-perturbative effects that come into play after multi-streaming.

The author would like to thank Rajaram Nityananda for his advice, encouragement and for many illuminating discussions.

This preprint was prepared with the AAS LaTeX macros v3.0.

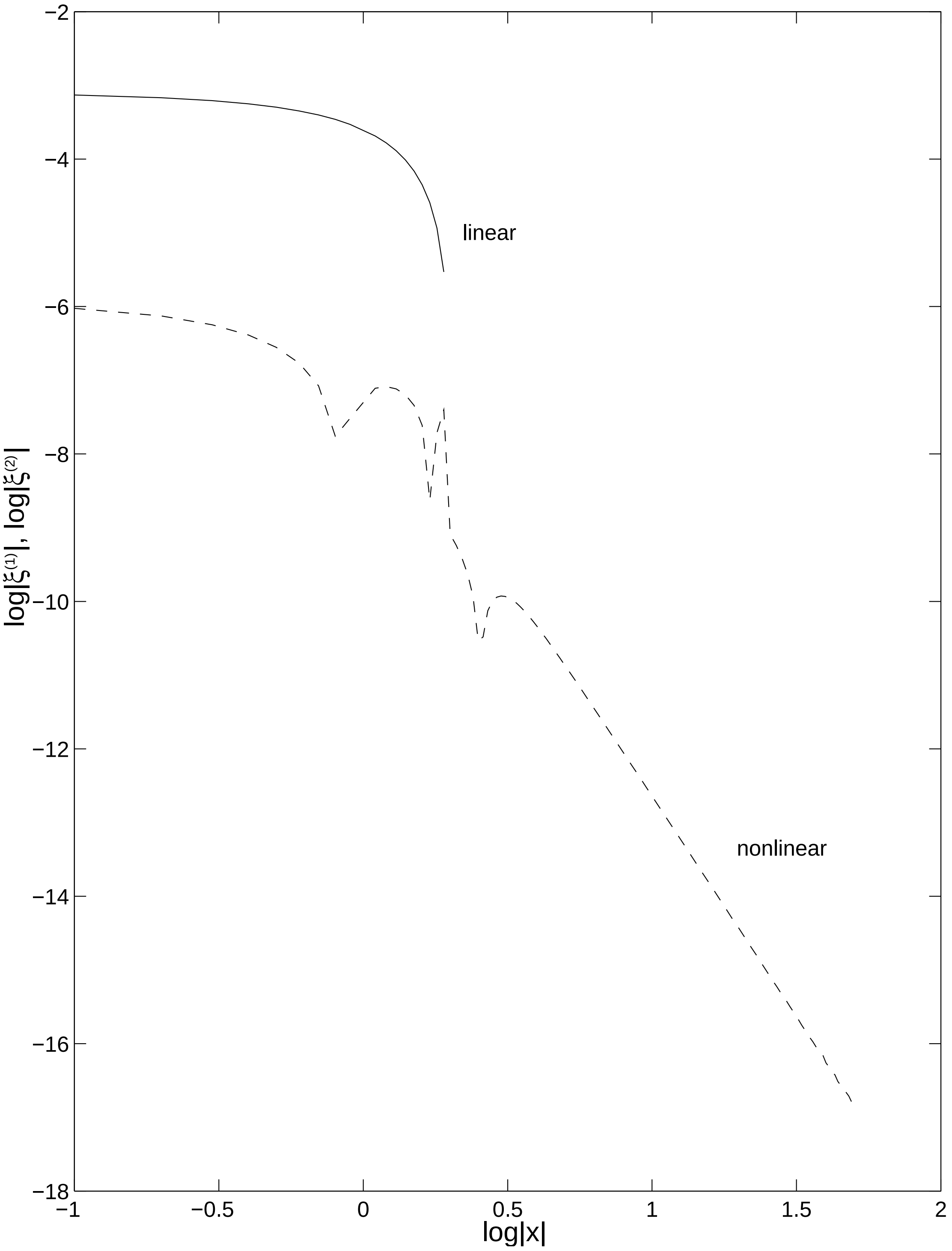

Fig. 1.— The linear two point correlation and its lowest order nonlinear correction for the case where the initial two point correlation correspnds to the self convolution of a sphere. The two curves have been given arbitrary displacements along the $y$ axis for convenience of displaying.

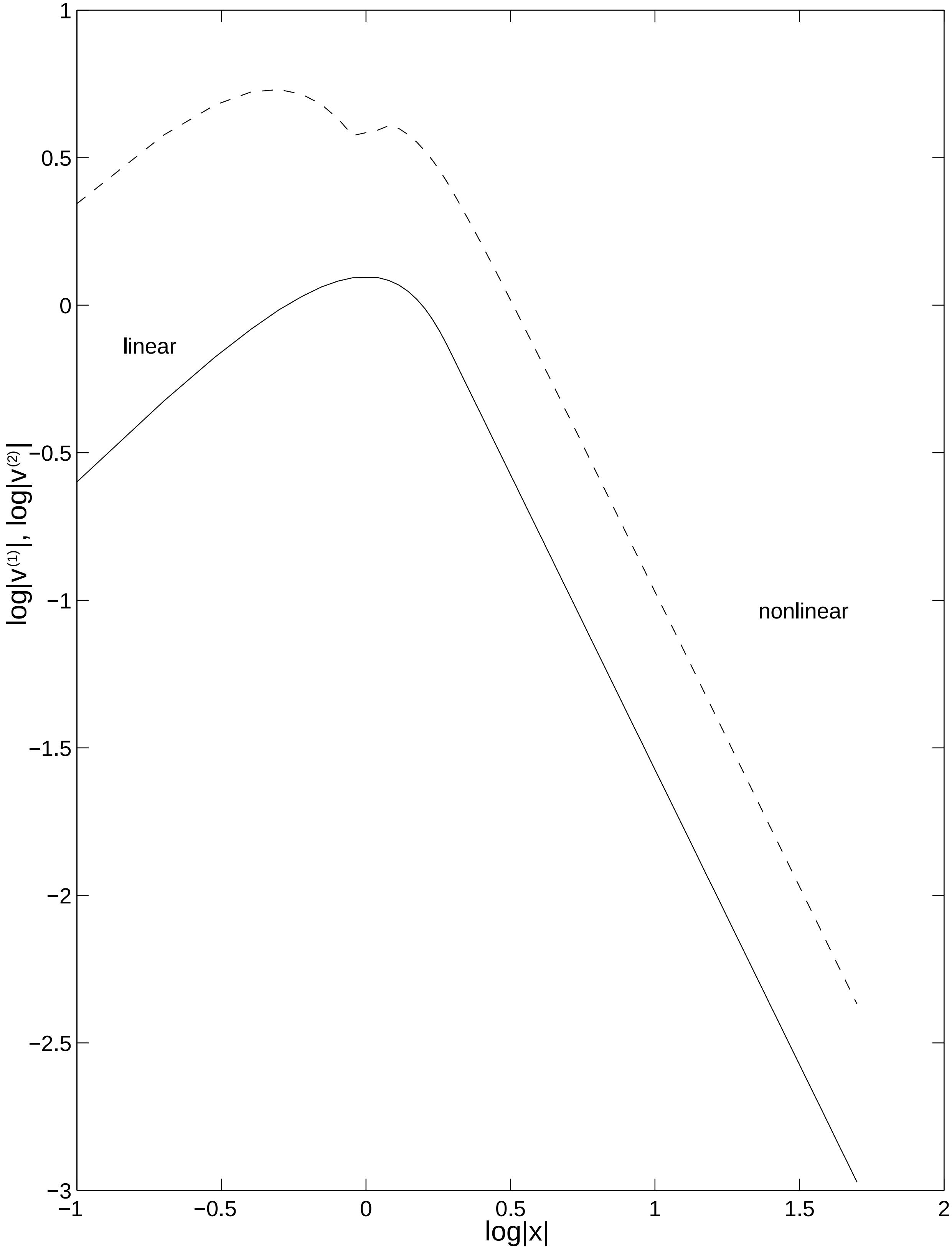

Fig. 2.— The linear pair velocity and its lowest order nonlinear correction for the case where the initial two point correlation correspnds to the self convolution of a sphere. The two curves have been given arbitrary displacements along the $y$ axis for convenience of displaying.

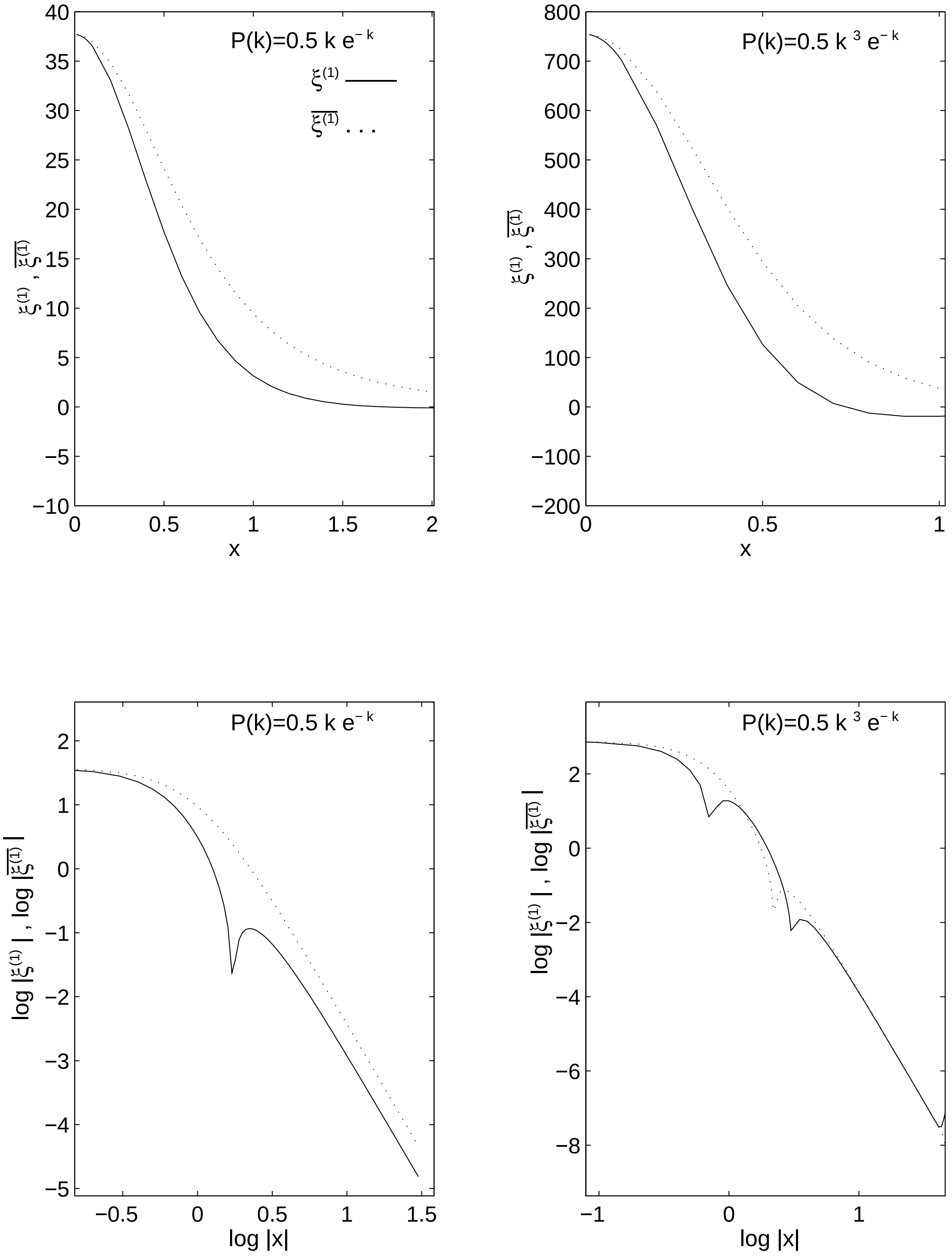

Fig. 3.— The initial two pint correlation $\xi^{(1)}(x)$ shown by the solid line and it average $\overline{\xi^{(1)}}(x)$ shown by the dotted line for two of the different initial power spectra considered.. The two plots on top show the small $x$ behaviour.

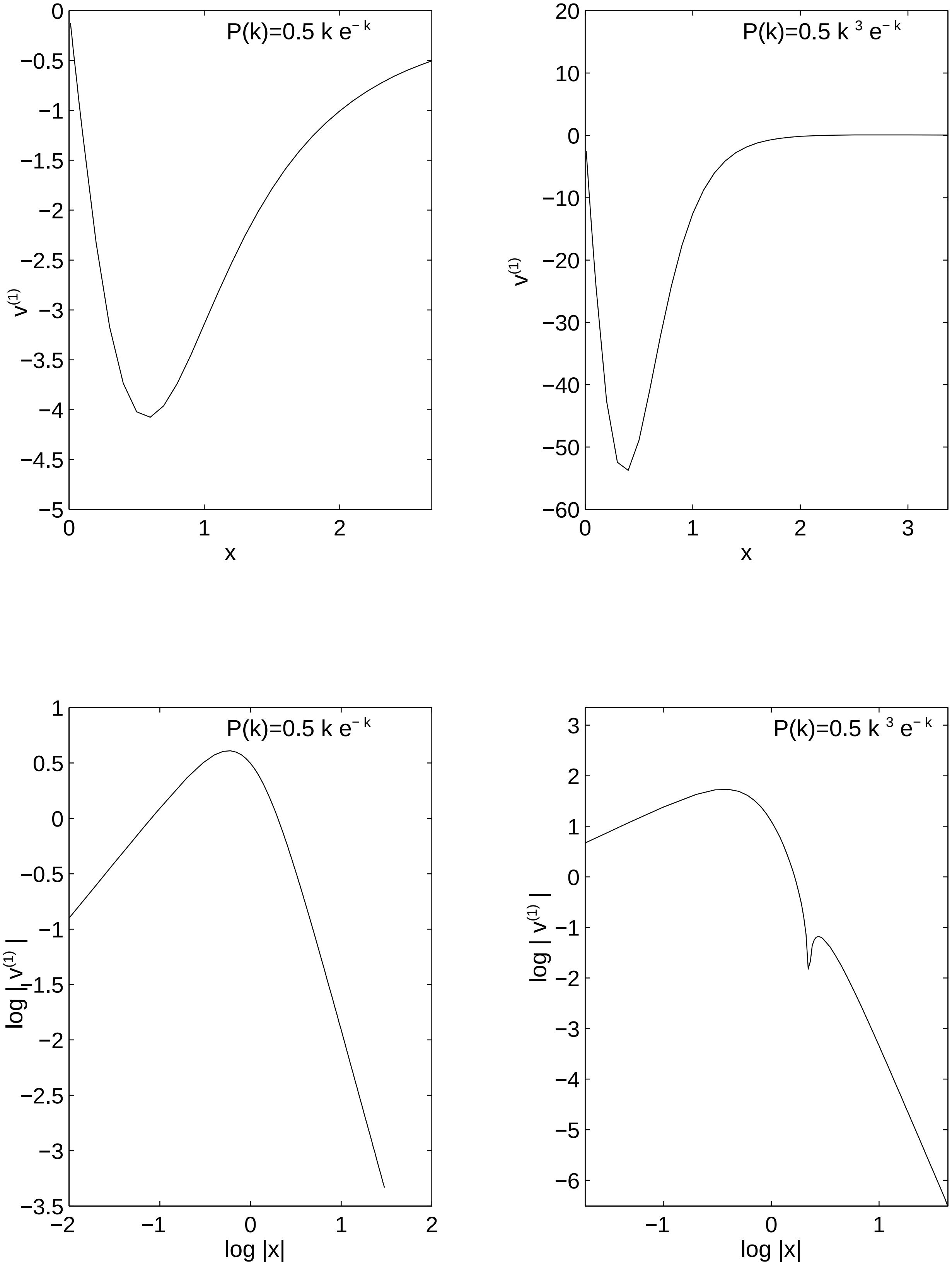

Fig. 4.— The initial pair velocity for two of the different initial power spectra considered. The two plots on top show the small $x$ behaviour.

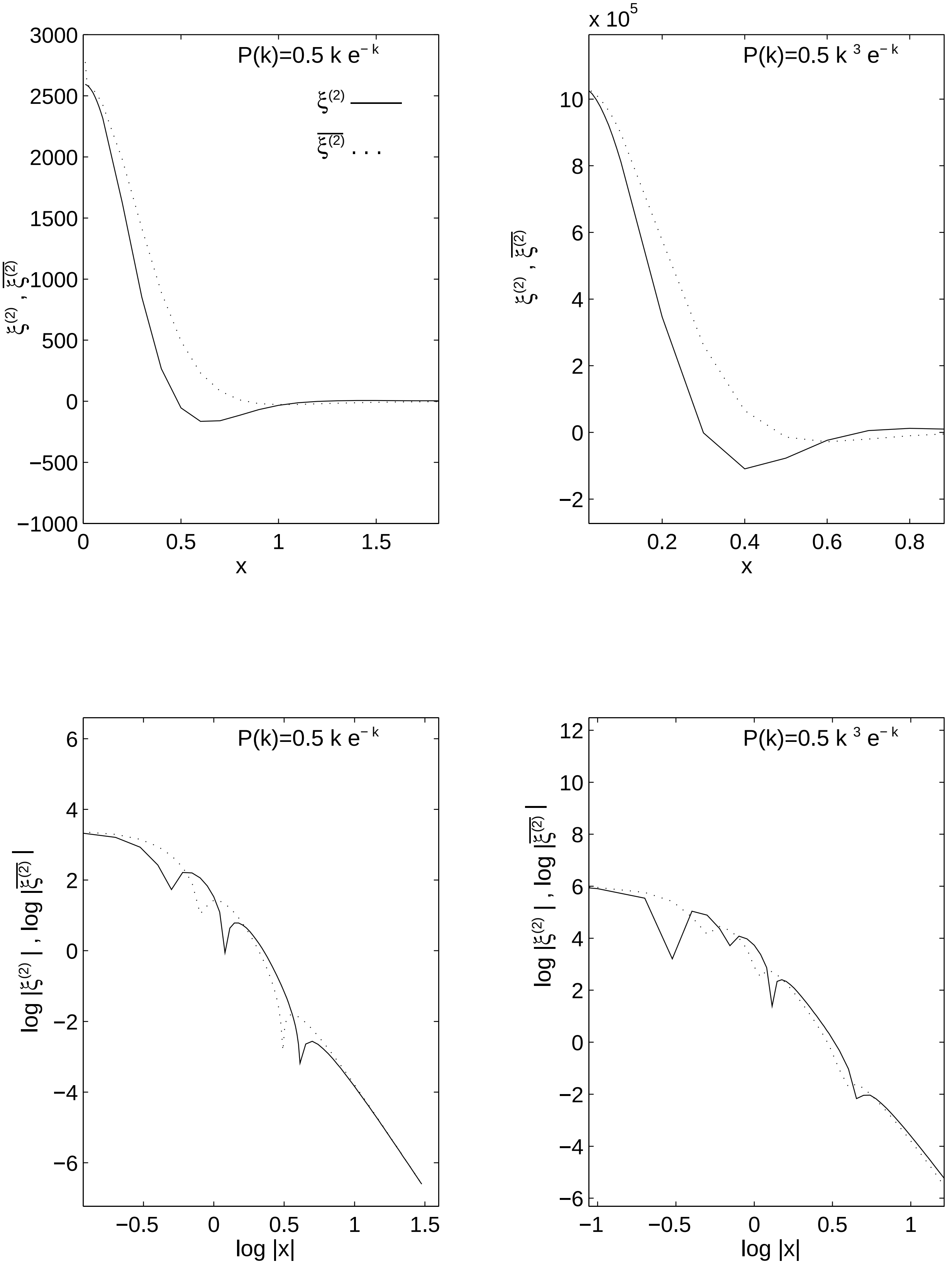

Fig. 5.— The nonlinear correction to the two pint correlation $\xi^{(2)}(x)$ shown by the solid line and it average $\overline{\xi^{(2)}}(x)$ shown by the dotted line for two of the different initial power spectra considered. The two plots on top show the small $x$ behaviour.

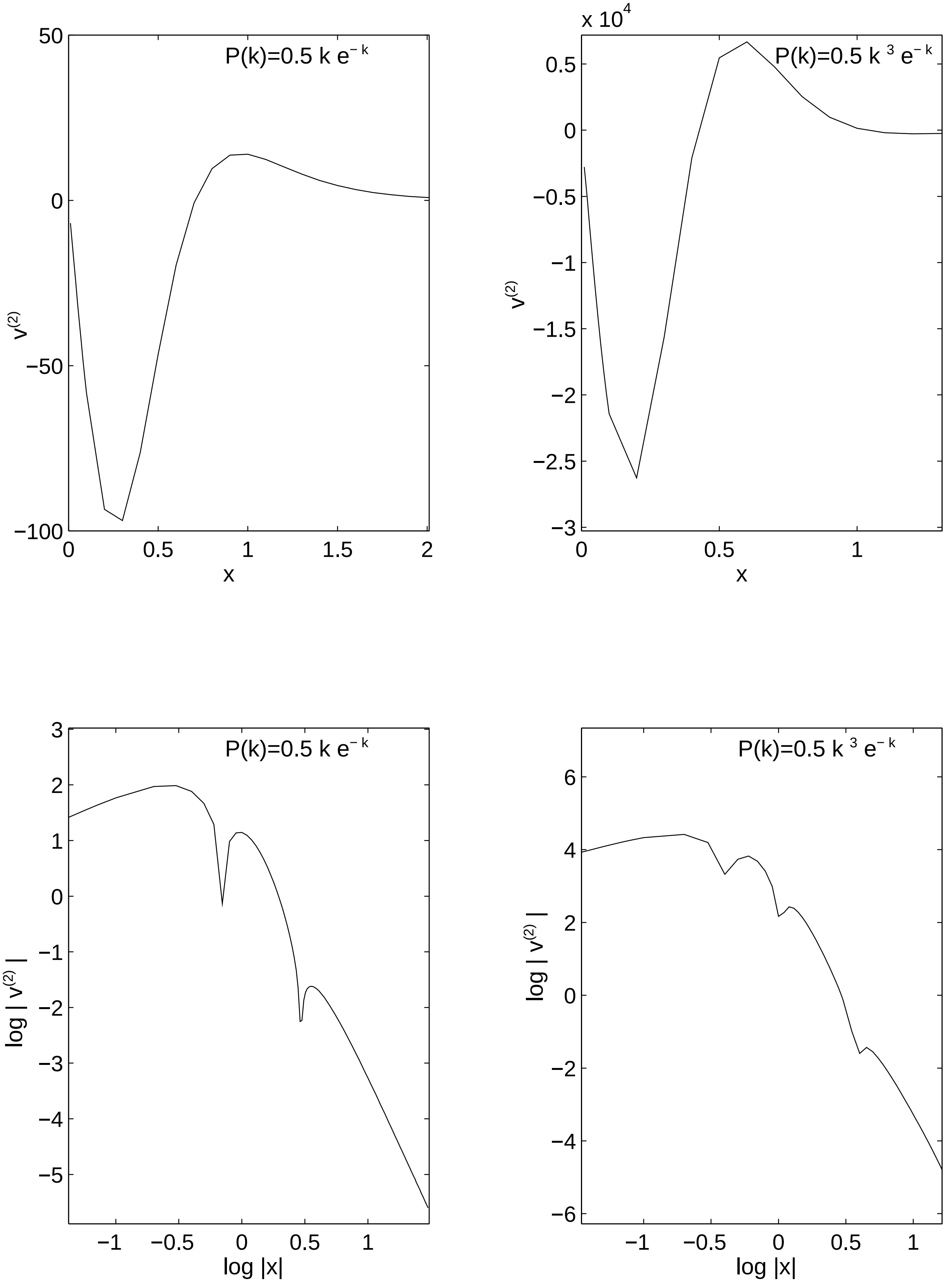

Fig. 6.— The nonlinear correction to the pair velocity for two of the different initial power spectra considered. The two plots on top show the small $x$ behaviour.

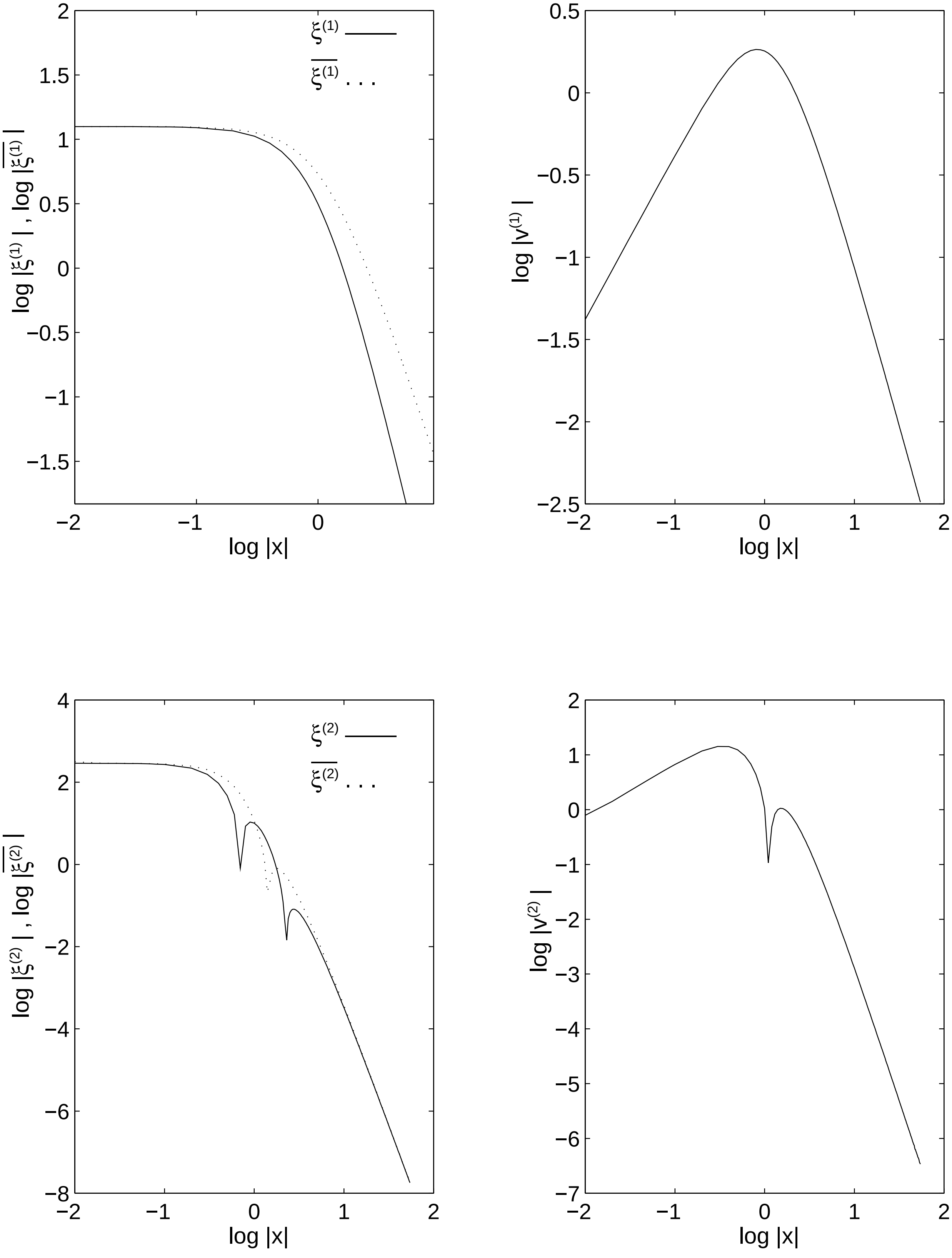

Fig. 7.— The linear two point correlation function $\xi^{(1)}(x)$, its average $\overline{\xi^{(1)}}(x)$ and the pair velocity $v^{(1)}(x)$ , and the repective nonlinear corrections $\xi^{(2)}(x)$,$\overline{\xi}(x,t)$ and $v^{(2)}(x)$ for the initial power spectrum $P(k) = e^{-k}$.

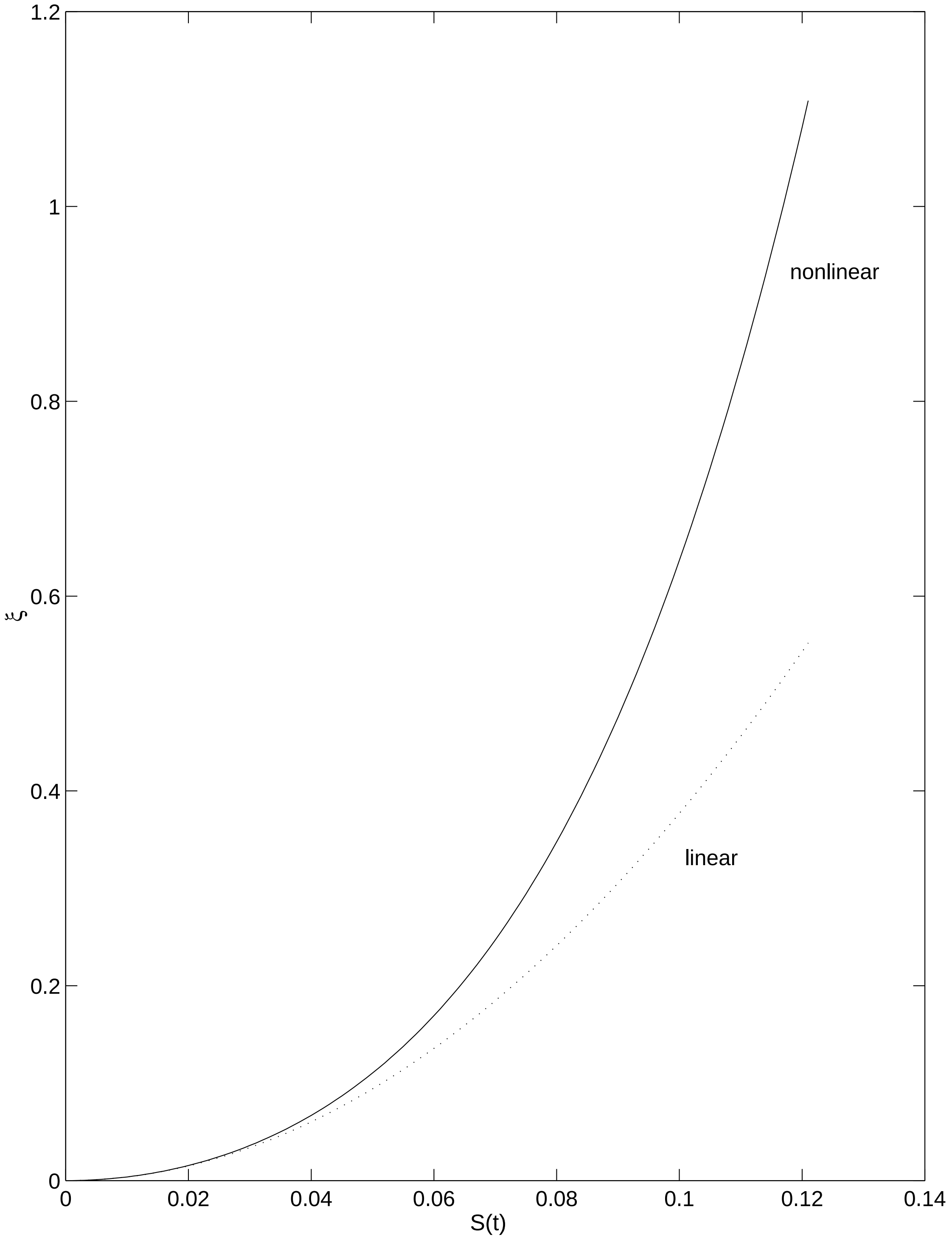

Fig. 8.— The solid line shows the two point correlation function $\xi(x,t) = \xi^{(1)}(x,t) + \xi^{(2)}(x,t)$ at $x = 0$ as a function of the scale factor $S(t)$. The dotted line shows just the linear contribution.

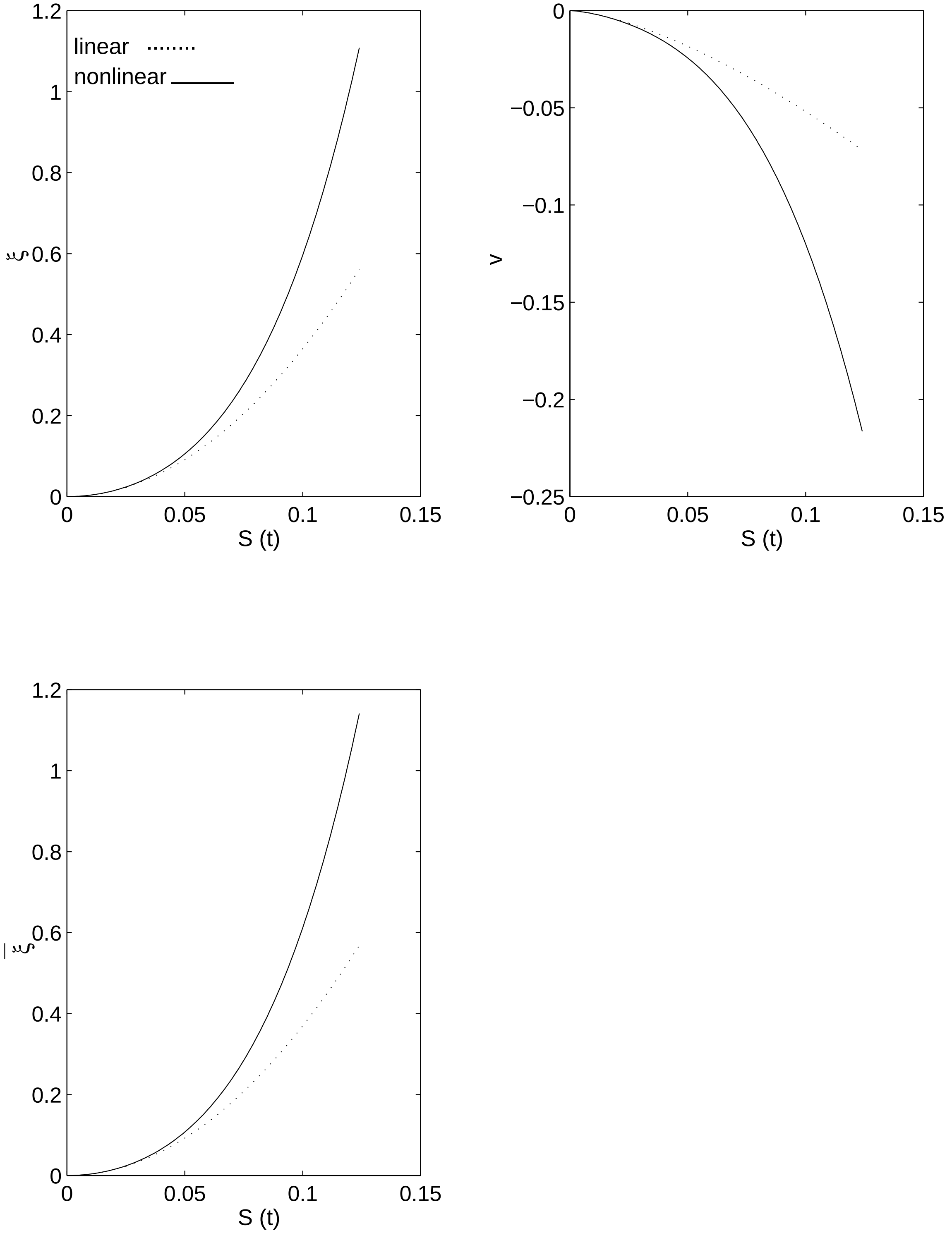

Fig. 9.— The solid curves show the two point correlation function $\xi(x,t) = \xi^{(1)}(x,t) + \xi^{(2)}(x,t)$, its average $\overline{\xi}(x,t)$ and the pair velocity $v(x,t)$ at $x = .1$ as a function of the scale factor $S(t)$. The dotted line shows just the linear contirbution to these quantities.

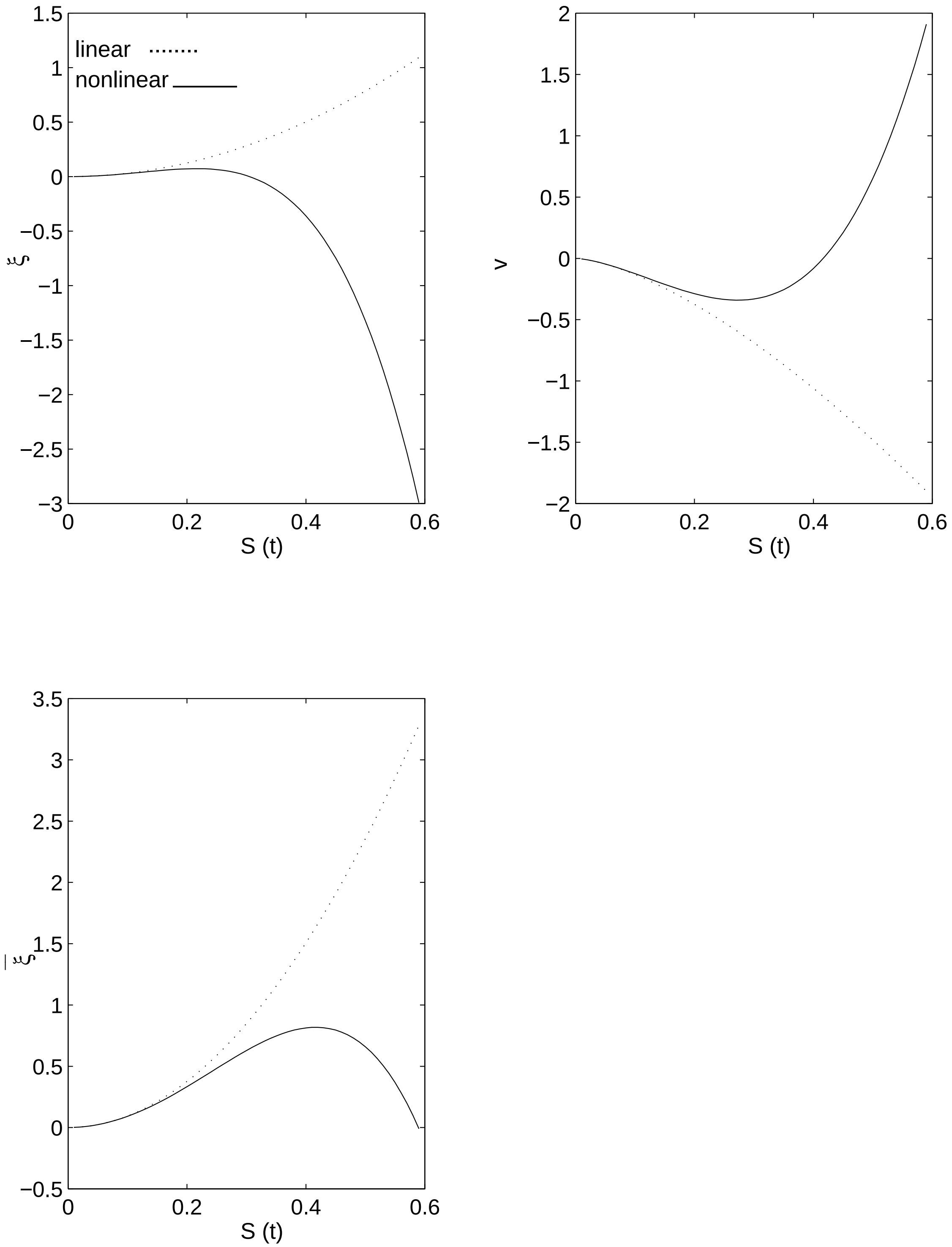

Fig. 10.— The solidcurves show the two point correlation function $\xi(x,t) = \xi^{(1)}(x,t) + \xi^{(2)}(x,t)$, its average $\overline{\xi}(x,t)$ and the pair velocity $v(x,t)$ at $x = 1.0$ as a function of the scale factor $S(t)$. The dotted line shows just the linear contirbution to these quantities.

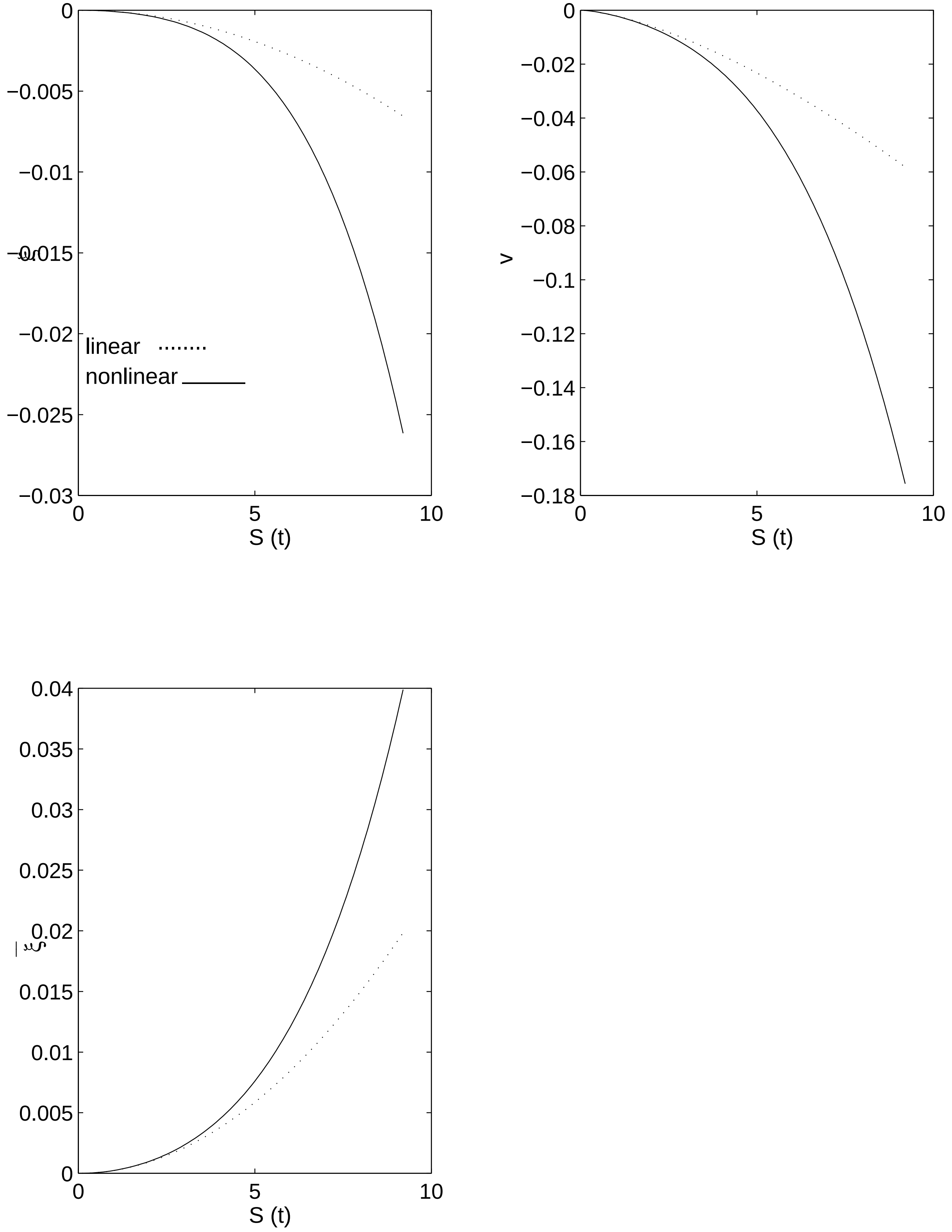

Fig. 11.— The solidcurves show the two point correlation function $\xi(x,t) = \xi^{(1)}(x,t) + \xi^{(2)}(x,t)$,its average $\bar{\xi}(x,t)$ and the pair velocity $v(x,t)$ at $x = 20.$ as a function of the scale factor $S(t)$. The dotted line shows just the linear contirbution to these quantities.

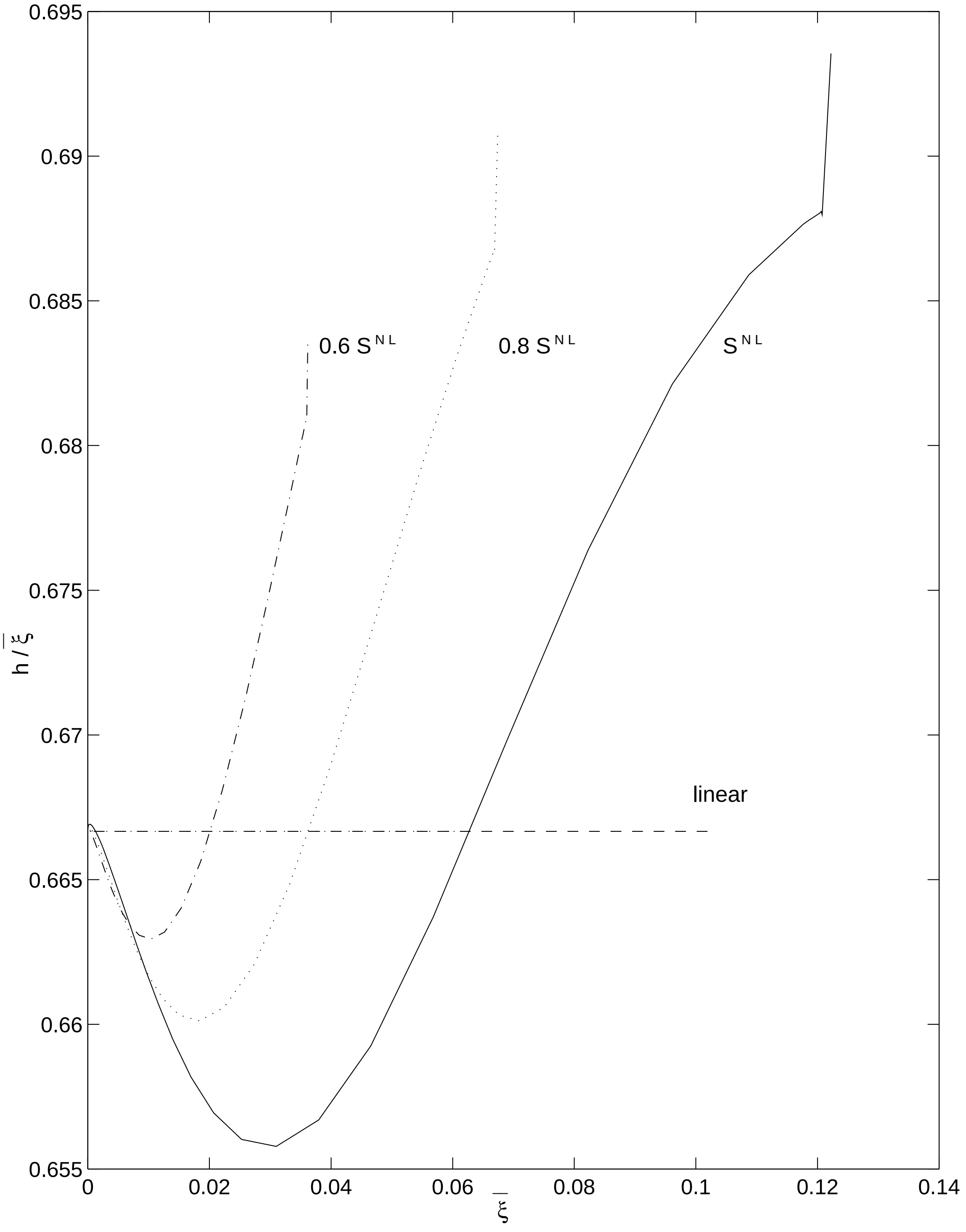

Fig. 12.— The ratio $\frac{h(x,t)}{\bar{\xi}(x,t)}$ for the initial power spectrum $P(k) = e^{-k}$ is shown as a function of $\bar{\xi}(x,t)$ for different values of the scale factor $S(t)$. $S^{NL}$ corresponds to the value of the scale factor for which the nonlinear correction to the two point correlation at $x = 0$ is one tenth of the linear quantity $(\xi^{(2)}(x,t) \sim 0.1\xi^{(1)}(x,t))$.

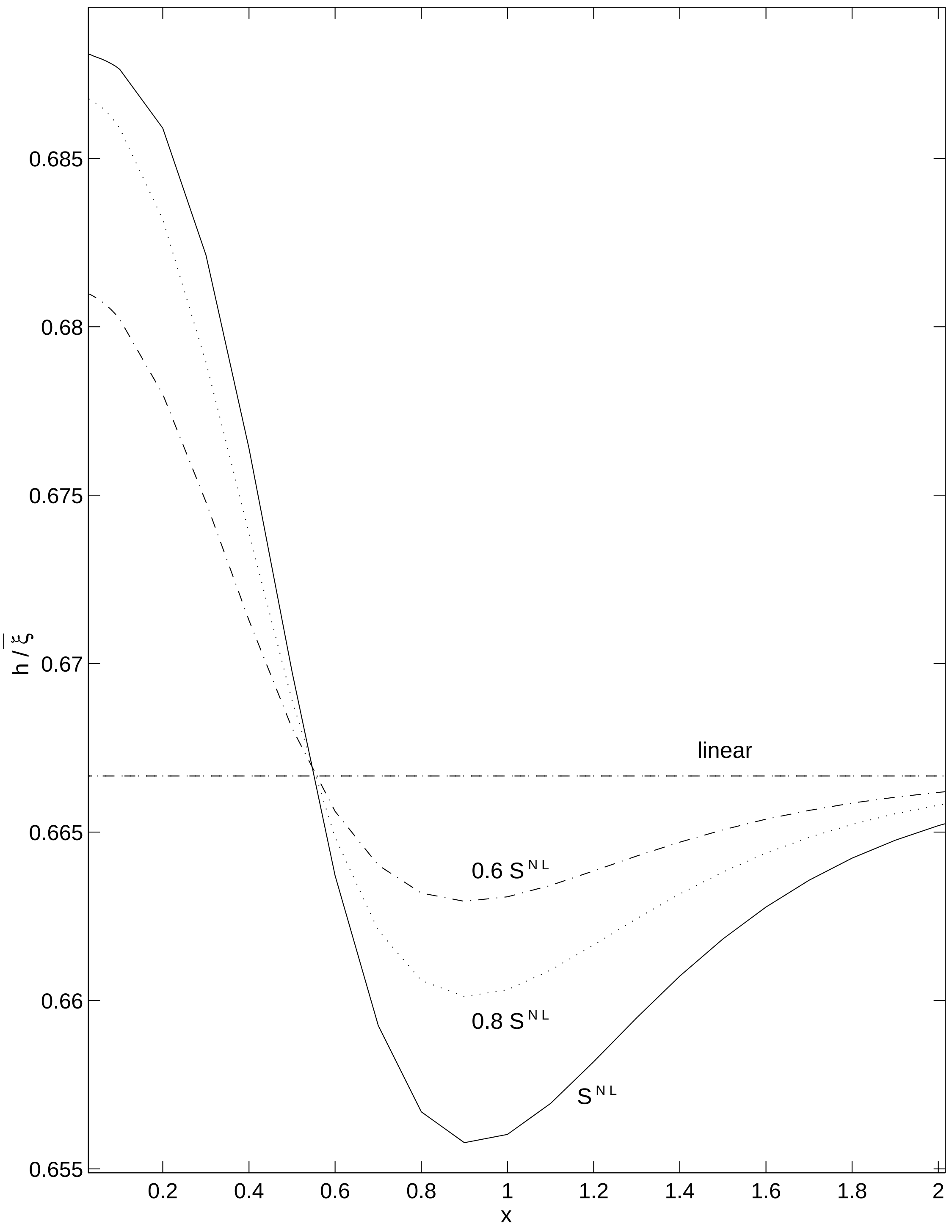

Fig. 13.— The ratio $\frac{h(x,t)}{\bar{\xi}(x,t)}$ for the initial power spectrum $P(k) = e^{-k}$ is shown as a function of $x$ for different values of the scale factor $S(t)$. $S^{NL}$ corresponds to the value of the scale factor for which the nonlinear correction to the two point correlation at $x = 0$ is one tenth of the linear quantity $(\xi^{(2)}(x,t) \sim 0.1\xi^{(1)}(x,t))$.